\documentclass[reprint,amsmath,amssymb,aps,pra]{revtex4-2}

\usepackage{float}
\usepackage{xcolor}
\usepackage{soul}
\usepackage{braket}
\usepackage{tikz}
\usetikzlibrary{arrows.meta}
\usepackage[caption=false]{subfig}
\usepackage{ragged2e}
\DeclareCaptionJustification{justified}{\justifying}
\usepackage[outdir=./]{epstopdf}

\usepackage{graphicx}
\usepackage{dcolumn}
\usepackage{bm}

\begin{document}

\title{A Hierarchy of Multipartite Nonlocality and Device-Independent Effect Witnesses}

\author{Peter Bierhorst} 
\email{plbierho@uno.edu}
\author{Jitendra Prakash}
\email{jprakash@uno.edu}

\affiliation{Mathematics Department, University of New Orleans, New Orleans, Louisiana 70148, USA}

\date{\today}

\begin{abstract}
According to recent new definitions, a multi-party behavior is \textit{genuinely multipartite nonlocal} (GMNL) if it cannot be modeled by measurements on an underlying network of bipartite-only nonlocal resources, possibly supplemented with local (classical) resources shared by all parties. The new definitions differ on whether to allow entangled measurements upon, and/or superquantum behaviors among, the underlying bipartite resources. Here, we categorize the full hierarchy of these new candidate definitions of GMNL in three-party quantum networks, highlighting the intimate link to device-independent witnesses of network effects. A key finding is the existence of a behavior in the simplest nontrivial multi-partite measurement scenario (3 parties, 2 measurement settings, and 2 outcomes) that cannot be simulated in a bipartite network prohibiting entangled measurements and superquantum resources -- thus witnessing the most general form of GMNL -- but can be simulated with bipartite-only quantum states \textit{with} an entangled measurement, indicating an approach to device independent certification of entangled measurements with fewer settings than in previous protocols. Surprisingly, we also find that this (3,2,2) behavior, as well as the others previously studied as device-independent witnesses of entangled measurements, can all be simulated at a higher echelon of the GMNL hierarchy that allows superquantum bipartite resources while still prohibiting entangled measurements. This poses a challenge to a theory-independent understanding of entangled measurements as an observable phenomenon distinct from bipartite nonlocality.
\end{abstract}

\maketitle

Quantum nonlocality~\cite{BELL} is a fascinating phenomenon that can be convincingly demonstrated in experiments of two spatially separated parties~\cite{hensen:2015,shalm:2015,giustina:2015,rosenfeld:2017}. Quantum mechanics also predicts nonlocal effects in experiments of three or more spatially separated parties. Naturally, a three-party experiment should only be considered \textit{genuinely multi-party nonlocal} if it exhibits some nonlocal behavior beyond the two-party type, ruling out scenarios where for instance two parties observe nonlocality with each other while the third party's statistics are not correlated with the first two in any way.

A first approach to defining genuine multipartite nonlocality (GMNL), introduced by Svetlichny~\cite{svetlichny:1987} and later refined~\cite{bancal:2013,gallego:2012}, proposes that a probability distribution of experimental outcomes be considered GMNL if it cannot be expressed as a convex mixture of distributions where each one factors into a product of at-most-bipartite nonlocal distributions. However, this definition admits anomalies~\cite{contreras:2021,schmid:2020,coiteux:2021}, for instance: If one measuring party simultaneously participates in two parallel but unrelated two-party Clauser-Horne-Shimony-Holt (CHSH~\cite{CHSH}) experiments, one with the second party and the other with the third party, the combined statistics of all three parties will be classified as GMNL according to Svetlichny-type definitions.

\begin{figure}
\includegraphics{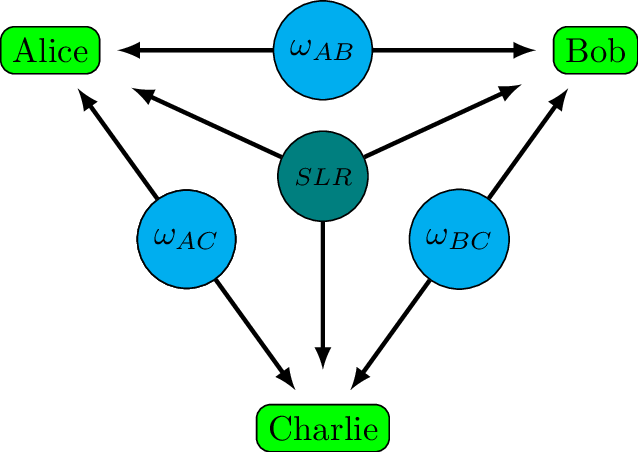}
\caption{\label{f:tripartite} A bipartite network model for a tripartite scenario. Tripartite behaviors that \textit{cannot} be induced by an underlying bipartite network model of the above form -- bipartite nonclassical sources ($\omega$) possibly supplemented with classical randomness shared by all three parties ($SLR$) -- are considered genuinely multipartite nonlocal (GMNL) according to recent new definitions~\cite{schmid:2020,bierhorst:2021,coiteux:2021}.}
\end{figure}

Recently, some authors~\cite{schmid:2020,bierhorst:2021,coiteux:2021} have proposed new definitions of GMNL based on whether a behavior can be simulated by an underlying network of bipartite nonlocal resources, possibly with access to local/classical resources shared by all parties (shared randomness). Fig.~\ref{f:tripartite} gives a schematic representation of such an underlying network for the three-party scenario, where a bipartite resource such as $\omega_{AB}$ shared by Alice and Bob could be an entangled Bell state $(\ket{00}+\ket{11})/\sqrt 2$, but three-way nonclassical states such as the GHZ state~\cite{GHZ} are disallowed. According to the new paradigm, a three-party behavior is considered GMNL if it \textit{cannot} be induced by an underlying network like that of Fig.~\ref{f:tripartite}. The parallel-CHSH-experiment example of the previous paragraph would here be ruled (only) bipartite nonlocal.

The new definitions~\cite{schmid:2020,bierhorst:2021,coiteux:2021} differ on the impositions made on the underlying network. The strictest of these definitions -- that is, the one which would categorize the \textit{largest} class of behaviors as (only) bipartite nonlocal -- is that of Coiteux-Roy, Wolfe, and Renou~\cite{coiteux:2021}. This definition allows the parties to perform entangled measurements, and also allows superquantum nonsignaling bipartite resources (such as Popescu-Rohrlich (PR) boxes~\cite{PRBOX}) in the underlying bipartite network. Tripartite behaviors ruling out this class, which can be achieved with appropriate measurements on the three-way entangled GHZ state~\cite{coiteux:2021,mao:2022}, naturally also rule out other definitions with more restrictions on the networks such as a definition disallowing entangled measurements~\cite{bierhorst:2021}, a definition disallowing superquantum bipartite resources~\cite{schmid:2020}, or a fourth candidate definition disallowing both. Recent experimental results~\cite{huang:2022,mao:2022,cao:2022} provide initial evidence, subject to fair-sampling-type assumptions, for the existence of three-party behaviors that cannot be modeled by even the most general underlying bipartite networks of Ref.~\cite{coiteux:2021}.
Different perspectives on what phenomena transcend that of (only) bipartite nonlocality motivate a closer study of the new definitions of GMNL that are less restrictive than that of Ref.~\cite{coiteux:2021}. To illustrate, observe that device-independent and self-testing witnesses of entangled measurements~\cite{rabello:2011,bancal:2018,renou:2018} are fundamentally multipartite phenomena, requiring at a minimum two distant parties and a third ``entangling'' party in between: any strictly two-party setup involving entangled measurements on different subsystems can always be easily simulated by a higher dimensional setup that does not employ entangled measurements (see Supplementary Material (SM) Section~1). Constraints derived under notions of GMNL that disallow entangled measurements will indeed be intimately linked to device-independent certificates of entangled measurements, a crucial tool for teleportation and entanglement swapping protocols in quantum networks~\cite{bennett:1993}. A device-independent perspective suggests disallowing superquantum nonsignaling bipartite resources (i.e., PR boxes~\cite{PRBOX}) among the $\omega$ sources in Fig.~\ref{f:tripartite} as nonphysical, but a more foundational perspective seeking a better theory-independent understanding of the nature of entangled measurements, which have recently been argued to remain poorly understood~\cite{gisin:2019}, recommends consideration of the GMNL paradigm where superquantum resources are allowed. We will consider both viewpoints.

In this letter, we study the full hierarchy of new definitions of GMNL and classify their interrelationships for the tripartite scenario. A main result of this work is the demonstration of a quantum behavior, using entangled measurements on bipartite-only quantum states, that witnesses the most general form of multi-party nonlocality -- that disallowing entangled measurements and superquantum resources in the Fig.~\ref{f:tripartite} network -- in the simplest possible (3,2,2) scenario of 3 measuring parties, 2 measurement settings per party, and 2 possible outcomes for each measurement. This behavior demonstrates an important separation between different definitions of GMNL, while also providing a promising approach to the task of device-independent certification of entangled measurements with the fewest-possible number of settings and outcomes -- reducing the number of settings from previous scenarios achieving this task \cite{rabello:2011,bancal:2018,renou:2018}. Note that as this behavior is not considered GMNL according to the stricter definition of \cite{coiteux:2021}, the nonfanout inflation technique \cite{wolfe:2019} used in \cite{coiteux:2021,mao:2022} is inapplicable for demonstrating the weaker notion of GMNL studied here, and our proof uses a different approach invoking self-testing \cite{supic:2020}.

This $(3, 2, 2)$ behavior demonstrates GMNL according to the definition where the $\omega$ in Fig.~\ref{f:tripartite} are limited to quantum-achievable resources. We continue the study by asking whether this behavior is still GMNL according to a paradigm in which superquantum resources (i.e., nonsignaling Popescu-Rohrlich (PR) boxes \cite{PRBOX}) are allowed for the underlying bipartite network, while still prohibiting entangled measurements and superquantum generalizations thereof. We find -- perhaps surprisingly -- that bipartite PR box networks can simulate the (3,2,2) behavior discussed above \textit{without} appealing to entangled measurements (or superquantum generalizations of the notion). Hence this behavior exhibits only bipartite nonlocality according to the GMNL definition allowing nonsignaling resources in Fig.~\ref{f:tripartite}. 

Motivated by this finding, we asked whether such a model exists for the more complicated behavior introduced by \cite{rabello:2011}, which has been studied in various forms~\cite{bancal:2018,renou:2018} as the canonical behavior certifying the presence of an entangled measurement in a fully device-independent manner. Similarly, we find a model for the behavior of Ref.~\cite{rabello:2011} using a network of bipartite PR boxes without entangled measurements. Hence none of these behaviors bear a theory-independent signature of the phenomenon of entangled measurements (i.e., without reference to the axioms of quantum mechanics), raising questions about exactly what such a signature might be, or if it exists.

We now give a precise formulation of the bipartite network model in which we rigorously derive our results. The three parties Alice, Bob and Charlie of Fig.~\ref{f:tripartite} make choices of measurements represented by respective random variables $X$, $Y$, and $Z$, and record measurement outcomes $A$, $B$, and $C$. An experiment is then characterized by the \textit{behavior} $P(A,B,C|X,Y,Z)$, the settings-conditional outcome distribution. Behaviors $P(ABC|XYZ)$ that can be induced by a network of the form in Fig.~\ref{f:tripartite} are said to be not GMNL, where the precise class of behaviors singled out differs based on the nature of the bipartite sources $\omega_{PQ}$ and the form of the measurements allowed to Alice, Bob, and Charlie.

\begin{figure}
\includegraphics[scale=0.8]{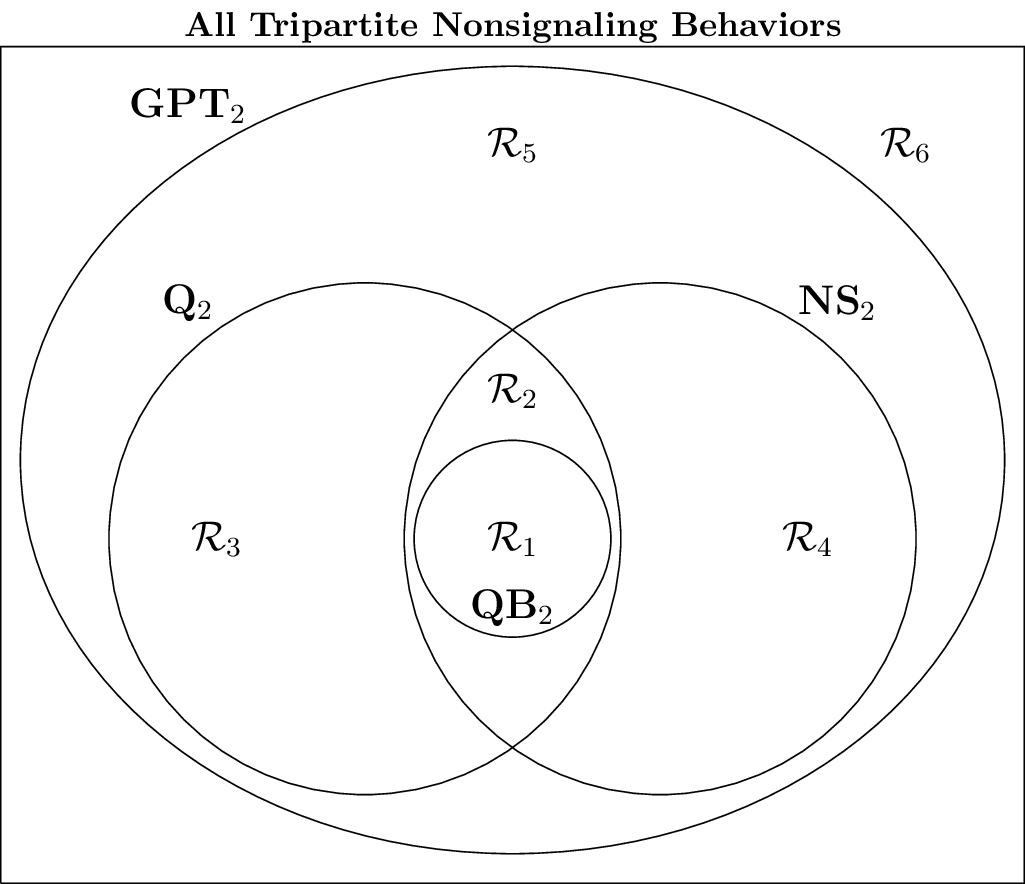}
\begin{tabular}{p{2.6cm}p{2.8cm}p{2.8cm}}
\textcolor{white}{spac}Behavior \textcolor{white}{space}Class & Superquantum\textcolor{white}{sp} Bipartite Sources  & Entangled Measurements \\
\hline
\multicolumn{1}{c}{$QB_2$} &\multicolumn{1}{c}{No}&\multicolumn{1}{c}{No} \\
\multicolumn{1}{c}{$NS_2$}&\multicolumn{1}{c}{Yes}&\multicolumn{1}{c}{No} \\
\multicolumn{1}{c}{$Q_2$} &\multicolumn{1}{c}{No}&\multicolumn{1}{c}{Yes}\\
\multicolumn{1}{c}{$GPT_2$}&\multicolumn{1}{c}{Yes}&\multicolumn{1}{c}{Yes} \\
\end{tabular}
\caption{\label{f:venn} Summary of features for the different models of an underlying network of bipartite-only systems in the tripartite scenario. The Venn diagram illustrates the containment  relationships for the corresponding classes of behaviors.}
\end{figure}

$QB_2$ is the smallest class of behaviors in the hierarchy of bipartite network models, which are summarized in Fig.~\ref{f:venn}. Here, the bipartite sources $\omega_{PQ}$ are taken to be quantum states $\rho_{PQ}$, so that the joint quantum state of the system is of the form $\rho_{AB}\otimes\rho_{BC}\otimes\rho_{AC}$. The parties Alice, Bob, and Charlie apply quantum measurements [positive operator-valued measures (POVMs)] to their respective systems, but must separately measure subsystems shared with different players. This is a scenario of ``quantum boxes" (motivating the choice of name $QB_2$), where quantum states are effectively input-output machines as entangling dynamics on the states are prohibited.
Because of this, the POVM elements of Bob (for example), which act on the state space of the reduced system $\text{tr}_{AC}[\rho_{AB}\otimes\rho_{BC}]$, are expressible in the separable form \cite[Proposition 6.5]{Watrous2018} \begin{equation}\label{e:separable}
\sum_i c_iR^A_i \otimes R^C_i,
\end{equation} where each $R^P_i$ is a rank-one projector acting on the portion of Bob's state shared with player $P$ and the $c_i$ are positive real constants not greater than one (SM 2). The class of separable measurements of form \eqref{e:separable} is in fact slightly larger than those measurements strictly admitting a quantum box description \cite{bennett:1999}.

The framework $QB_2$ disallows superquantum resources for the $\omega_{PQ}$ in Fig.~\ref{f:tripartite}, which can be justified on practical grounds: superquantum correlations such as those of the PR boxes are generally expected to be nonphysical, and in the device-independent certification perspective the validity and completeness of quantum mechanics is generally assumed. Quantum behaviors outside of $QB_2$ require either entangled measurements or three-way entangled states, and so device-independently witness the presence of at least one of these resources.

If one accepts the position that only quantum resources should be considered for the bipartite resources in Fig.~\ref{f:tripartite}, but that entangled measurements should be permitted, one arrives at the larger class of bipartite network behaviors $Q_2$. This corresponds to the notion of GMNL given in Definition 2 of Ref.~\cite{schmid:2020}. $Q_2$ is precisely the boundary for a behavior exhibiting tripartite entangled states device-independently; any tripartite quantum behavior lying outside this set certifies the presence of a three-way-entangled quantum state (in particular, a \textit{genuinely network 3-entangled} state as defined in~\cite{navascues:2020}).

Another option for extending the class $QB_2$ is to allow for superquantum resources such as PR boxes \cite{PRBOX} while instead maintaining the prohibition on entangled measurements. For the observable phenomenon of (bipartite) nonlocality, the most abstract definition of this phenomenon -- that without any appeal to the axioms of quantum mechanics -- involves black boxes that can violate Bell inequalities while respecting the no-signaling conditions. The framework $NS_2$ allows the classical manipulation whereby outputs of some of the bipartite boxes are used as inputs to other bipartite boxes, expanding the scope of simulable tripartite behaviors~\cite{barrett:2005}. Finally, the largest class $GPT_2$ (standing for generalized probabilistic theories) allows for both superquantum bipartite sources and entangled measurements (and possibly superquantum generalizations thereof); this corresponds to the GMNL definition of Ref.~\cite{coiteux:2021}.

The containment relationships of the four sets are summarized in Fig.~\ref{f:venn}. It is known that some of the containments are strict: region $\mathcal R_4$ can be seen to be nonempty due to the presence of PR box correlations in $NS_2$ while Tsirelson's bound~\cite{tsirelson:1993} rules these out of $Q_2$, and the results of~\cite{coiteux:2021,mao:2022} demonstrate quantum behaviors in region $\mathcal R_6$. It is conjectured in Section V.C of Ref.~\cite{coiteux:2021a} that there are correlations outside $NS_2$ but inside $GPT_2$, but to date we are unaware of an argument definitively proving the existence of behaviors in either region $\mathcal R_3$ or $\mathcal R_5$.

The three-party behavior introduced by~\cite{rabello:2011} and further studied by~\cite{bancal:2018,renou:2018} as a device-independent certificate of entangled measurements can, due to this certifying property, be situated in the current context as lying outside $QB_2$ but inside $Q_2$; see Proposition 7 in Ref.~\cite{coiteux:2021a} for an extended discussion. (Whether these behaviors are in region $\mathcal R_2$ or $\mathcal R_3$ requires further analysis; we answer this question later.) The behavior of~\cite{rabello:2011} and all of its later-studied variants are characterized by having more than two setting choices for at least one of the parties. In contrast, the following result shows that a behavior in $Q_2\setminus QB_2$ can be found for the simplest possible (3,2,2) scenario. All behaviors in any simpler measurement scenario can always be simulated with bipartite resources and shared local randomness (SM 3).

\medskip

\noindent \textit{Theorem 1}. There is a behavior $P(ABC|XYZ)$ in $Q_2$ with binary input and output random variables satisfying the conditions $P(B=0|Y=1)>0$, $P_{Y=1,B=0}(AC|XZ)$ maximally violates the CHSH inequality, and ${P(A=B|X=0, Y=0)=1}$. Furthermore, no behavior in $QB_2$ can satisfy these conditions.

\medskip

\noindent The behavior in $Q_2$ is obtained as follows: Alice and Bob, and Bob and Charlie, each share a Bell pair ${\ket {\Phi^+} = (\ket{00}+\ket{11})/\sqrt 2}$. No Alice-Charlie state is used. On setting $Y=1$, Bob performs an (entangled) Bell state measurement on his two portions of the Bell pairs; conditioned on observing the outcome corresponding to $\ket{\Phi^+}$, which occurs with probability 1/4, Bob reports outcome $B=0$ and all other Bell measurement outcomes are binned into outcome $B=1$. Conditioned on $B=0$, Alice and Charlie possess $\ket{\Phi^+}$ on which they can perform measurements maximally violating the CHSH inequality with Alice measuring $\sigma_z$ on setting $X=0$. On setting $Y=0$, Bob measures (only) his qubit shared with Alice in the same direction $\sigma_z$, which ensures $P(A=B|X=0,Y=0)=1$. The exact behavior $P(ABC|XYZ)$ is \begin{align*}
&\frac{1+(-1)^{A\oplus B}\delta_{X,0}}{8} &&\text{ if } Y = 0, \\
&\frac{\delta_{B,1}}{4} + \frac{(-1)^B}{4}\mathrm{CHSH}(AC|XZ) &&\text{ if } Y = 1,
\end{align*} where $\mathrm{CHSH}(AC|XZ) = (2+(-1)^{A\oplus C \oplus XZ}\sqrt{2})/8$.

That such behaviors cannot exist in $QB_2$ follows by the following intuition, which we make precise and prove in SM Section 4: Assume Bob can make only a separable measurement on setting $Y=1$ (the proof does not assume separability of any other measurement). This measurement cannot create new entanglement between Alice and Charlie, but Alice and Charlie must be measuring an entangled Bell state to maximally violate CHSH, and so this must be a Bell state they initially possess via $\omega_{AC}$. Then since Bob is not entangled with $\omega_{AC}$, from his perspective Alice is measuring a fully mixed state and it will be impossible for him to do any better than blind guessing when trying to align his outcome with Alice's for setting $Y=0$.

As in \cite{rabello:2011}, our rigorous proof relies crucially on self-testing, but we encounter a notable complication in the need to link a conditional \textit{post}-Bob-measurement CHSH violation to restrictions on Bob's ability to align with Alice's outcome when he chooses a different measurement setting, requiring a new argument that necessarily cedes improved (but not perfect) prospects for Bob to align outcomes with Alice. Our proof applies in full generality, i.e., assuming only POVMs (see SM Section 6, where we borrow an argument from \cite{PeresQT} instead of the standard one \cite{NC} for POVM-to-projective-measurement dilation.) on potentially mixed states, and while we do assume a maximal violation of the CHSH inequality, this leads to a robust upper bound (strictly less than 1) on $P(A=B|X=0,Y=0)$ such that robustness results for self-testing \cite{kaniewski:2016} provide a clear approach for lifting the argument to experimentally testable constraints, and thereby a device-independent witness of entangled measurements in the simplest possible (3,2,2) scenario.

We extend our analysis by asking whether this behavior lies in region $\mathcal R_2$ or $\mathcal R_3$. One might be tempted to think that the (3,2,2) behavior described above cannot be simulated in $NS_2$, due to the well known prohibition on ``nonlocality swapping"~\cite{short:2006,barrett:2007}. However:

\medskip

\noindent \textit{Theorem 2}. There exists a behavior in $NS_2$ meeting the conditions of Theorem 1.

\medskip

\begin{figure}[ht]
\centering
\subfloat[]{
\includegraphics[scale=0.55]{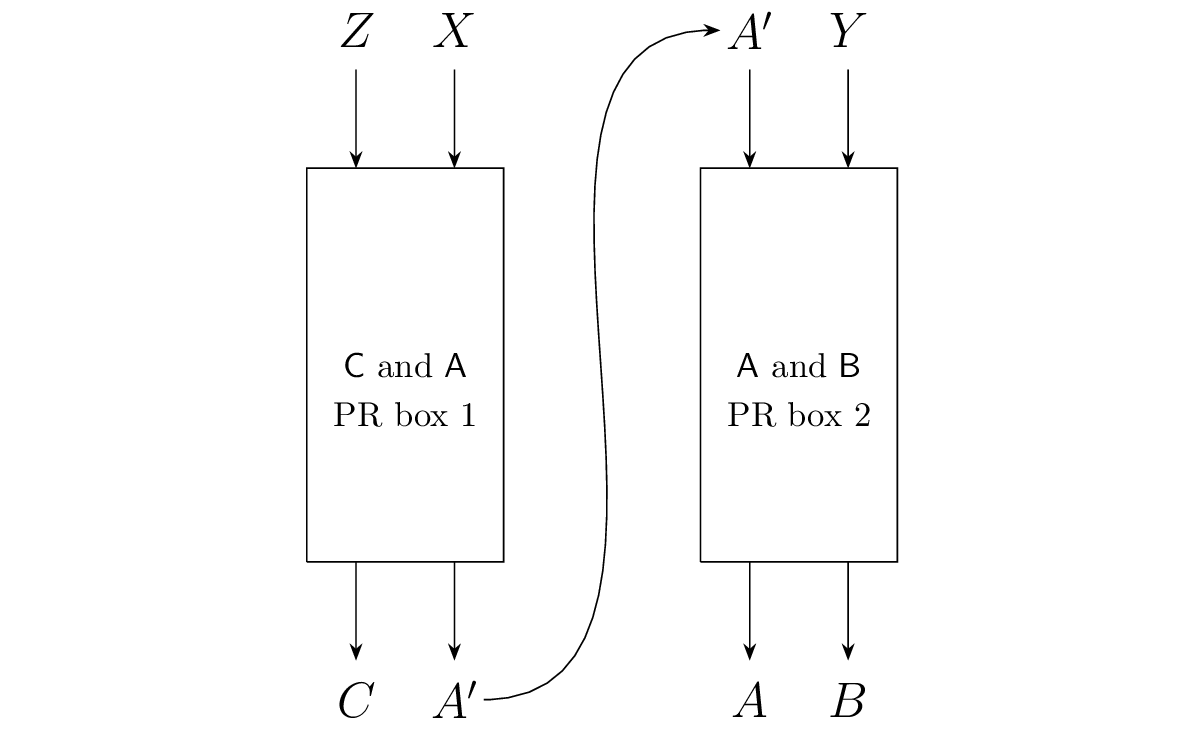}}\\
\subfloat[]{\includegraphics[scale=0.55]{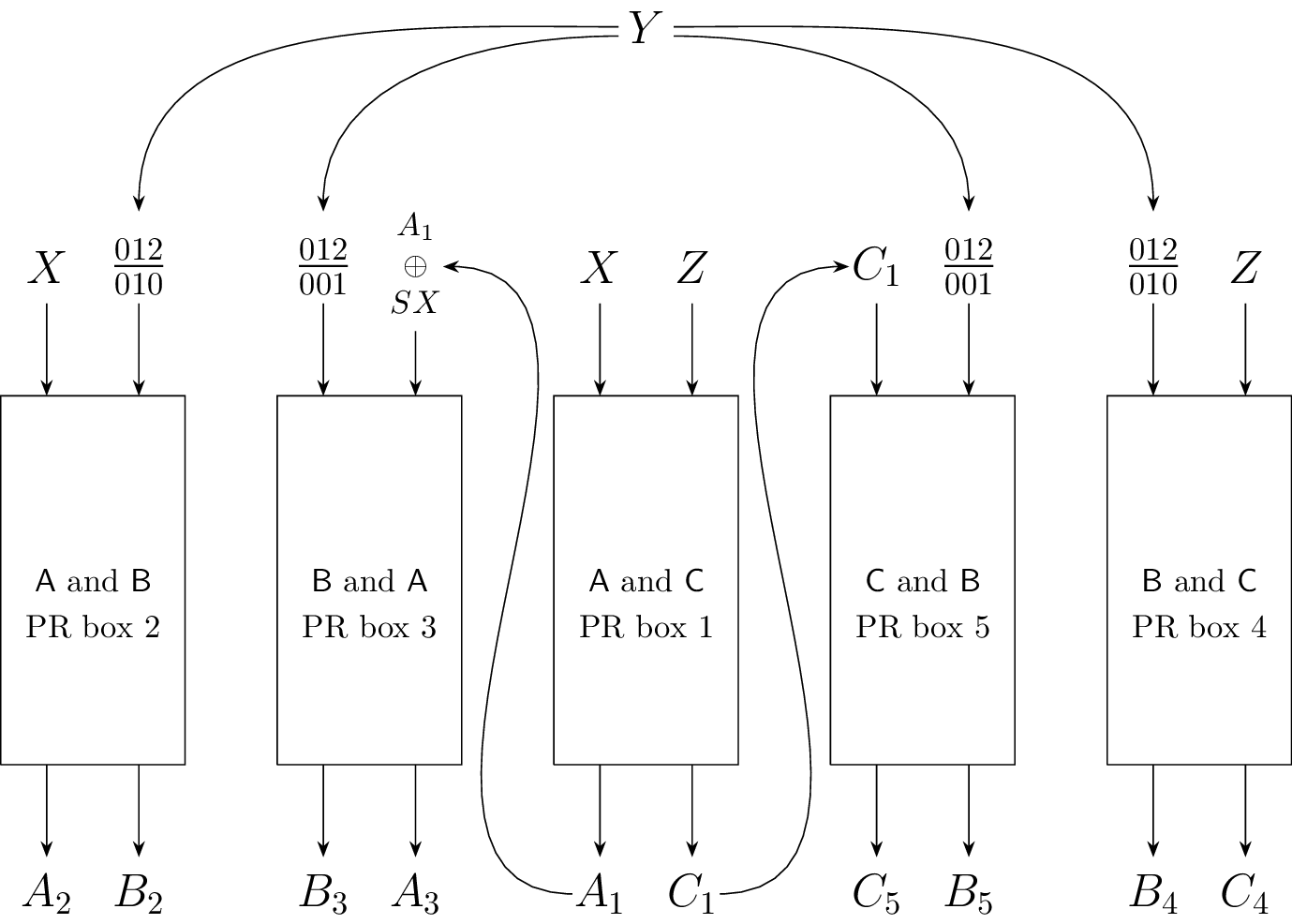}}
\caption{Each rectangle in the figures above denotes a PR box. The PR box behavior is the unique bipartite behavior satisfying $A\oplus B = XY$ and $P(A|X) = P(B|Y) = 1/2$ for binary inputs $X,Y$ (denoted with in-flowing arrows above) and outputs $A,B$ (out-flowing arrows). Text inside the box, for example, in the left-most box of part (a), ``$\mathsf{C}$ and $\mathsf{A}$'' means that the box is shared by Charlie and Alice, with input $Z$ and output $C$ of Charlie on the left part of the box and input $X$ and output $A^{\prime}$ of Alice on the right part of the box. (a) A network of PR boxes spoofing entangled measurements by witnessing the claims of Theorem~2. (b) A network of PR boxes spoofing entangled measurements by witnessing the claims of Theorem~3. Alice's outcome is $A=A_2\oplus A_3$, Charlie's outcome is $C=C_4\oplus C_5$, and Bob's outcome is $(B_A, B_C)=(B_2\oplus B_3, B_4\oplus B_5)$ for $Y\in\{0,1\}$ and $(B_A, B_C)=(S, B_2\oplus B_3\oplus B_4\oplus B_5)$ for $Y=2$. $S$ is a fully random binary bit shared by Alice and Bob, which can be implemented with a sixth PR box with constant inputs 0. The fraction $\frac{012}{010}$ above PR box 2 means that if Bob receives $Y = 0$ (resp., $1$ and $2$), he inputs $0$ (resp., $1$ and $0$) in that box. A similar convention holds for the other boxes.}
\label{f:prnets}
\end{figure}

\noindent \textit{Proof}. Fig.~\ref{f:prnets}(a) gives an example of a PR box network that results in the behavior $P(ABC|XYZ)$ given by \begin{align*}
&\frac{\delta_{A, B}}{4} &&\text{ if } Y = 0, \\
&\frac{\delta_{B,1}}{4} + \frac{(-1)^{B}}{2}\mathrm{PR}(AC|XZ) &&\text{ if } Y = 1,
\end{align*} where $\mathrm{PR}(AC|XZ) = \frac{\delta_{A\oplus C, XZ}}{2}$. This behavior satisfies the conditions of Theorem 1 with the modification that $P_{Y=1,B=0}(AC|XZ)$ violates the CHSH inequality beyond the Tsirelson's bound to the nonsignaling maximum of 4. A convex mixture of this behavior with classical behaviors can induce violations of the CHSH inequality to only the quantum maximum of $2\sqrt 2$.

\medskip

A possible idea for why the (3,2,2) behavior might fail to bear a theory-independent signature of an entangled measurement is that tripartite quantum behaviors outside $QB_2$ only signify \textit{either} the presence of entangled measurements \textit{or} three-way entangled sources, and it is only with the additional assumption of the absence of three-way entangled sources that the (3,2,2) behavior certifies entangled measurements specifically. Indeed, Jordan's lemma ensures that any (3,2,2) behavior can be simulated with (non-entangled) measurements on qubits~\cite{masanes:2005} and we provide in SM Section 5 an explicit example satisfying the conditions of Theorem 1 with a GHZ state. The assumption of the absence of three-way entangled sources is also required in~\cite{bancal:2018,renou:2018} for noise-robust device-independent certification of entangled measurements, and while the assumption can be well motivated physically in appropriate setups, it can be argued to technically represent a weakening to a semi-device-dependent scenario. However, this assumption is \textit{not} invoked in the original argument concerning the noise-free behavior of Ref.~\cite{rabello:2011}. But we find that even the original behavior of~\cite{rabello:2011} is simulable with networks of bipartite PR boxes. 

In the scenario of~\cite{rabello:2011}, re-formulated as a Bell game, Alice and Charlie still have binary settings and outcomes, but now Bob has three settings $Y\in \{0,1,2\}$, each with four outcomes modeled as a binary pair $B=(B_A, B_C)$. When $Y\in\{0,1\}$, two subgames are won if $A\oplus B_A=XY$ and $C\oplus B_C = ZY$; these are two parallel CHSH games played by Alice-Bob and Bob-Charlie. When $Y=2$, the winning condition is $A\oplus C=XZ\oplus (XB_A\oplus B_C)$; this constitutes four variants of an Alice-Charlie CHSH game, corresponding to each potential value of $B$. As argued in \cite{rabello:2011}, a strategy utilizing bipartite Bell states and a Bell basis measurement for $Y=2$ can win all the CHSH games to the quantum maximum ($\cos^2(\pi/8)\approx 85\%$) whereas no strategy without an entangled measurement can do so even if tripartite entangled states are available. However, a network of PR boxes as in Fig.~\ref{f:prnets}(b) fulfills the following theorem:

\medskip

\noindent \textit{Theorem 3}. The Bell game of Rabello et al.~\cite{rabello:2011} described above can be won with probability 1 by a behavior in $NS_2$.

\medskip

The results of this paper provide a minimally complex approach to witnessing entangled measurements, situated in the wider context of classifying different notions of genuine multipartite nonlocality. The techniques of Theorem 1 may also be useful in other paradigms: for example, in the triangle network \textit{without} global shared randomness, the behavior of \cite{renou:2019} was conjectured to require entangled measurements but was only recently proven to do so \cite{sekatski:2022}. And the results of Theorem 2 and 3 indicate that claims of non-simulability by PR boxes for behaviors invoking entangled measurements (such as is suggested for the behavior in \cite{renou:2019} but this remains unproven) must be carefully evaluated. Whether any tripartite behaviors exist in  region $\mathcal R_3$ of Fig.~\ref{f:venn} remains an open question with important implications for a theory-independent understanding of entangled measurements.

\begin{acknowledgments}
The authors acknowledge Elie Wolfe for stimulating conversations leading to consideration of the (3,2,2) behavior studied here, and for helpful comments on the manuscript. This work was partially supported by NSF Award No.~2210399 and AFOSR Award No.~FA9550-20-1-0067.
\end{acknowledgments}


%

\newpage
\onecolumngrid
\appendix*
\section{}

\subsection{Two party behaviors never require entangled measurements}\label{app:EM-not-required}

If two parties Alice and Bob share multiple quantum states, and perform entangled measurements on their different sub-states, it is always possible to re-write the shared quantum states as a single quantum state $\sum_{ij}\gamma_{ij}\ket i_A \otimes \ket j_B$ using bases $\{\ket i\}$ and $\{\ket j\}$ that ignore any subsystem structure. Then the corresponding measurements will no longer have any entangled structure. For example, suppose Alice and Bob possess two separate Bell pairs \begin{equation*}
(1/\sqrt 2)\left(\ket{0_A0_B}+\ket {1_A1_B}\right)\otimes(1/\sqrt 2)\left(\ket{0_{A'}0_{B'}}+\ket {1_{A'}1_{B'}}\right)
\end{equation*} and Alice makes an entangled measurement on her $A$ and $A'$ subsystems. Then using the relabeling $\ket 0=\ket{0_A}\otimes \ket{0_{A'}}$, $\ket 1=\ket{0_A}\otimes \ket{1_{A'}}$, $\ket 2=\ket{1_A}\otimes \ket{0_{A'}}$, $\ket 3=\ket{1_A}\otimes \ket{1_{A'}}$, Alice's measurements are replaced by (unentangled) measurements on a single 4-dimensional qudit.

\subsection{Separable measurements encompass ``quantum box" dynamics}\label{app:QB-dynamics}

We illustrate how Expression (1) in the main text encompasses ``quantum box" dynamics. Such dynamics include strategies where Bob measures his portion of $\rho_{AB}$, then depending on this outcome chooses a measurement on his portion of $\rho_{BC}$, possibly repeating this process multiple times across $\rho_{AB}$ and $\rho_{BC}$ if these consist of parallel sub-resources, with Bob finally reporting a final outcome $B$ as a function of these sub-measurements' outcomes.

First we state the following elementary fact (for a proof see Proposition 6.5 in \cite{Watrous2018}):

\medskip

\noindent\textbf{Fact.} Any operator of the form $\Pi = \sum_{i} P^A_i \otimes P^C_i$, where the $P^A_i$ and $P^C_i$ are positive operators, can be expressed in the form $\sum_j c_jR^A_j \otimes R^C_j$ where $R^A_j,R^C_j$ are rank one projectors and the $c_j$ are positive constants. If $\Pi$ is also a POVM element then the $c_j$ are bounded above by 1.

\medskip

Now let us consider a simple cascaded submeasurement approach performed by Alice on the subsystems she shares respectively with Bob and Charlie. Suppose Alice measures the register $A^{(b)}$, which is her portion of the state shared with Bob, with POVM elements $\{P^i \}_{i\in\{0,1\}}$; if her outcome to this measurement is 0, she measures her subsystem $A^{(c)}$ shared with Charlie with $\{S^i\}_{i\in\{0,1\}}$, but if her outcome is 1, she chooses a different measurement $\{T^i\}_{i\in\{0,1\}}$ for the Charlie subsystem. Then this whole measurement procedure can be represented as a single (separable) measurement with four measurement operators: \begin{equation*}
P^0_{A^{(b)}}\otimes S^0_{A^{(c)}} \quad P^0_{A^{(b)}}\otimes S^1_{A^{(c)}} \quad P^1_{A^{(b)}}\otimes T^0_{A^{(c)}} \quad P^1_{A^{(b)}}\otimes T^1_{A^{(c)}}.
\end{equation*} If she chooses to bin some of these together for a coarse-grained final outcome $A$, she can achieve the same behavior by employing a POVM whose corresponding element is the sum of the binned elements. Then by the above fact, this is representable with the form of Expression (1) in the main text.

A more complicated scenario with multiple subsystems comprising both $A^{(b)}$ and $A^{(c)}$, with various earlier measurements controlling choices of later measurements in the different subsystems, still leads to separable measurements where each measurement operator is of a form like \begin{equation*}
\left[ P_{A_1^{(b)}}\otimes Q_{A_2^{(b)}} \otimes R_{A_3^{(b)}}\right]_{A^{(b)}} \otimes \left[ S_{A_1^{(c)}}\otimes T_{A_2^{(c)}} \otimes U_{A_3^{(c)}}\right]_{A^{(c)}}
\end{equation*} in which each bracketed term is itself a projector, consisting of a tensor product of projectors. This is an example of a LOCC (local operations and classical communication) transformation performed by Alice upon her separate subsystems, with the ``classical communication" being transmitted between Alice's subsystems. Ref.~\cite{bennett:1999} shows that there are separable measurements that cannot be modeled with such a LOCC approach.

\subsection{$(3,2,2)$ is the simplest nontrivial scenario in the presence of shared local randomness}\label{app:3-2-2-simplest}

The allowance for three-way classical resources in Fig.~1 prevents certain evidently classical behaviors from being classified as GMNL, such as the fixed setting behavior in which Alice, Bob, and Charlie either all observe ``0" or all observe ``1" with equal probability. This behavior cannot be simulated with bipartite nonclassical systems alone; see Example 1 of Ref.~\cite{wolfe:2019} and the further discussion in Section V.D therein. It is still possible to study three-party networks under an assumption of the absence of three-way shared randomness, and with this assumption one can witness the presence of nonclassical effects in scenarios simpler than $(3,2,2)$: for example,~\cite{renou:2019} observes a form of nonlocality in a three-party network where each party has only one choice of setting. However, this is not possible in our paradigm: the $(3,2,2)$ scenario is the simplest scenario in which a nonsignaling behavior can rule out simulation by bipartite-only nonclassical sources and three-way shared local randomness.

To see why, first observe that we clearly cannot reduce the number of parties below three -- this results in a bipartite scenario, and so is of course bipartite simulable.

Furthermore, in an $n$-partite experiment witnessing GMNL each party must have at least two settings; if not, an experiment of $(n-1)$ parties would have the same ability to witness incompatibility with an underlying bipartite network. To illustrate, suppose one of the parties in an $n$-party experiment has only one setting; say Alice. Consider Alice to measure first, obtaining an outcome $A=a_i$ which occurs with probability $p(i)$, and let $P_i$ denote the behavior of the remaining players conditioned on the occurrence of this Alice outcome. In the three party version, Alice's lack of setting and the no-signaling property allows the following factorization of the behavior: ${P(ABC|YZ)=P(BC|YZ,A)P(A|YZ)=P(BC|YZ,A)P(A)}$, and $P_i=P(BC|YZ, A=a_i)$. An analogous factorization holds for a higher number of parties. The factorization shows us that if each $P_i$ is bipartite simulable as an $(n-1)$-partite behavior, we can simulate the $n$-party behavior with the following scheme: distribute networks capable of simulating each of the $P_i$ behaviors to the the other $(n-1)$ players, and distribute a classical random variable $\Lambda$ to all $n$ parties that takes the value $a_i$ with probability $p(i)$. During the experiment, Alice reports the value of the $\Lambda$ variable as her outcome $A$ while the remaining parties use the $P_i$-generating network whose whose index $i$ corresponds to the observed value of $\Lambda$. This will simulate the original behavior, and so the original $n$-party behavior is incompatible with an underlying bipartite network only if one of the $a_i$-conditional behaviors of the $(n-1)$ non-Alice parties is so incompatible.

Finally, every setting must have at least two outcomes. This is because a behavior with a one-outcome measurement setting will always be simulable by an underlying bipartite network if the reduced behavior without this setting choice is so simulable. This follows from the no-signaling principle: suppose (say) Alice only has one outcome $\mathcal O$ for her last measurement setting $m$. Then consider the scenario where Alice only has $(m-1)$ settings and no setting $m$. If the corresponding reduced behavior is bipartite simulable, then we can simulate the original behavior adding back in the $m$th setting as follows: when the $m$th setting is queried, Alice does the same thing as for one of her other $(m-1)$ measurement settings but just relabels all outcomes to the single possible outcome. Since the marginal distribution of the remaining parties is required to be the same regardless of Alice's setting by the no-signaling principle, the full original behavior is recovered this way. Mathematically, this corresponds to the equalities $P(A=\mathcal O, BC|X=m, YZ) = P(BC|X=m, YZ) =P(BC|X=m-1, YZ) =\sum_iP(A=a_i, BC|X=m-1, YZ)$.

By the above considerations, any scenario witnessing GMNL is always as or more complicated than a simplified scenario witnessing GMNL with $n\ge 3$ parties, at least two measurement settings per party, and at least two outcomes per measurement, and the $(3,2,2)$ scenario is minimally complicated among such scenarios. The above considerations also show that no scenario less complicated than $(3,2,2)$ can witness GMNL.

\subsection{Proof of Theorem 1}\label{app:Thm-1-proof}
As outlined in the main text, we employ self-testing \cite{supic:2020} combined with the separable structure of the measurements to demonstrate that behaviors in $QB_2$ cannot meet the conditions of Theorem 1. We remark this is different from the approaches of \cite{coiteux:2021,mao:2022,cao:2022} which all use the nonfanout inflation technique \cite{wolfe:2019} to derive constraints on the behavior class that we call $GPT_2$. That technique is inapplicable for demonstrating the weaker notion of GMNL studied here, because behaviors in $Q_2$ -- which can meet the conditions of Theorem 1 -- are not considered GMNL according to the stricter definition of \cite{coiteux:2021} that considers anything in $GPT_2$ bipartite-only nonlocal. Hence constraints on $GPT_2$ obtained with the nonfanout inflation technique, while necessarily obeyed by behaviors in $QB_2$, will never be incompatible with meeting the conditions of Theorem 1.

We restate the conditions of Theorem 1 as follows:
\begin{align}
&P(B=0|Y=1)>0 \text{ and }P_{Y=1,B=0}(AC|XZ) \text{ maximally violates CHSH inequality}\label{e:cond1}\\
&P(A=B|X=0, Y=0)=1\label{e:cond2}
\end{align}
Self-testing was also invoked in the arguments of Refs.~\cite{rabello:2011,bancal:2018,renou:2018}, but in these works, the CHSH violation restricting the structure of the state was not conditional on a third player's outcome as in \eqref{e:cond1} above. We thus require an additional argument to link this post-outcome CHSH violation to constraints on the pre-measurement Alice-Bob state that prevent condition \eqref{e:cond2} from being met. One consequence of this conditionality is that Bob's degree of failure of \eqref{e:cond2} is linked to $P(B=0|Y=1)$; quantitatively, we find below that $P(A\ne B|X=0,Y=0)$ is bounded below by half of $P(B=0|Y=1)$. A quantitative lower bound on $P(A\ne B|X=0,Y=0)$ such as this one, rather than just the impossibility of unit probability in \eqref{e:cond2}, is important in consideration of future work extending the following argument to a robust testable version featuring sub-maximal CHSH violations.

We prove Theorem 1 in the most general case of POVM measurements on mixed states. Some of the arguments below, however, apply only to projective measurements and/or pure states. In particular, self-testing results are generally formulated with an assumption of projective measurements on pure states; see Appendix B of Ref.~\cite{supic:2020} for a discussion of this assumption and its implications in the self-testing context. This sometimes-implicit assumption introduces subtleties for applying self-testing results to other contexts. Thus, for our theorem to hold in full generality, we adopt a strategy of first showing that if a $QB_2$ behavior meets the conditions \eqref{e:cond1}-\eqref{e:cond2}, this implies the existence of a (possibly different) behavior meeting the conditions of \eqref{e:cond1}-\eqref{e:cond2} for which the measured state is pure, some of the measurements are projective, and some of the separable measurement restrictions of $QB_2$ are still met. Then we show that this \textit{different} behavior leads to a contradiction; that is, no behavior meeting these modified restrictions can actually satisfy conditions \eqref{e:cond1}-\eqref{e:cond2}. This strategy requires some care since standard state purification and measurement dilation arguments do not necessarily preserve characteristic structures of $QB_2$ (separable measurements and a product structure of the measured states).

In the proof, we use the following properties of partial trace:

\medskip

\noindent \textbf{Fact.} For the tensor product Hilbert space $\mathcal H_X \otimes \mathcal H_Y$, the partial trace $\text{Tr}_Y$ has the following properties:
\begin{align}
\textnormal{Linearity: } & \text{Tr}_Y \left(\sum_i\lambda_i \rho^i_{XY}\right)=\sum_i\lambda_i \text{Tr}_Y \left(\rho^i_{XY}\right)\label{e:ptlinear}\\
\textnormal{Partial Cyclicity: } & \text{Tr}_Y \left[(I_X \otimes M_Y )\rho_{XY}\right]=\text{Tr}_Y \left[\rho_{XY}(I_X \otimes M_Y )\right]\label{e:ptcyclic}
\end{align}
where $\rho_{XY}$, $I_X$ (identity) and $M_Y$ operate on $\mathcal H_X\otimes \mathcal H_Y$, $\mathcal H_X$, and $\mathcal H_Y$ respectively
\medskip

\noindent\textit{Proof of Theorem 1}. An example of a behavior in $Q_2$ meeting conditions \eqref{e:cond1}-\eqref{e:cond2} is provided in the main text. We show here that no behavior in $QB_2$ can meet these conditions, employing a proof by contradiction.

\medskip

\noindent\textbf{Step 1: Simplifying states and measurements.} Assume that a behavior in $QB_2$ meets the conditions \eqref{e:cond1}-\eqref{e:cond2} with POVMs on a mixed state. Such a behavior implies the existence of a (possibly different) behavior in $QB_2$ meeting the conditions using the same POVMs on a pure state by the following convexity argument: by the nature of $QB_2$ the measured mixed state is of the form $\rho_{AB}\otimes\rho_{BC}\otimes\rho_{AC}$. One can represent each of the three component mixed states $\rho_{PQ} $ as a convex mixture of pure states $\sum\lambda_i\ket{\psi^{PQ}_i}\bra{\psi^{PQ}_i}$. Then applying the POVMs to the composite pure state \begin{equation*}
\left(\ket{\psi^{AB}_i}\bra{\psi^{AB}_i}\right)\otimes\left(\ket{\psi^{BC}_j}\bra{\psi^{BC}_j}\right)\otimes\left(\ket{\psi^{AC}_k}\bra{\psi^{AC}_k}\right)
\end{equation*} yields a behavior such that the convex mixture of all such behaviors with respective weights $\lambda_i\lambda_j\lambda_k$ recovers the original behavior. Clearly \eqref{e:cond2} must hold for each individual behavior in this convex mixture. Moreover, if a convex mixture of quantum behaviors satisfies the condition \eqref{e:cond1}, then at least some of the individual behaviors must satisfy this condition as well, since some of the individual behaviors must satisfy $P(B=0|Y=1)>0$ and the average CHSH value over all such behaviors is $2\sqrt 2$ requiring each individual behavior in this class to achieve $2\sqrt 2$ as CHSH values exceeding $2\sqrt 2$ are impossible. So some of the behaviors in the convex mixture meet the conditions \eqref{e:cond1}-\eqref{e:cond2} with POVMs on a pure state.

Thus it suffices to demonstrate impossibility of satisfying the conditions \eqref{e:cond1}-\eqref{e:cond2} in $QB_2$ with a pure state. To apply our argument, we require a further simplification of Alice and Charlie's measurements to be projective; this enables a direct application of self-testing results as well as some other simplifications. Replacing POVMs with projective measurements yielding the same behavior is always possible, but we are careful to employ a method that preserves the factored form of the state
\begin{equation}\label{e:state}
\ket{\psi}=\ket{\psi}_{AB}\otimes \ket{\psi}_{BC}\otimes \ket{\psi}_{AC}.
\end{equation}
The method described on p.~95-6 of \cite{NC} cannot be used because it involves the party that is replacing the POVM with a projective measurement to apply a unitary to the state which could entangle the two portions that the party shares with the two other parties. We instead follow a method close to that of \cite{PeresQT}. As we show in Theorem~3 of Section~\ref{app:POVM-dilation}, this approach allows us to replicate the behavior while replacing the state $\ket \psi$ with a new state $\ket \psi'$ of the form
\begin{align}\label{eq:psi-prime}
   \ket{\psi}' &= \ket \alpha_A\otimes\ket{\psi}_{AB}\otimes \ket{\psi}_{BC}\otimes \ket{\psi}_{AC} \otimes \ket \alpha_C\\
   &= \ket{\psi}_{AB}'\otimes \ket{\psi}_{BC}\otimes \ket{\psi}_{AC}'
\end{align}
where Alice performs a projective measurement on her previous state space plus an introduced qudit $\ket\alpha_A$, Charlie does similarly with $\ket\alpha_C$, and Bob's POVM is unchanged. Collecting the introduced qudits into respective states $\ket{\psi}_{AB}'$ and $\ket{\psi}_{AC}'$ then makes the state still conform with the $QB_2$-style factorization of \eqref{e:state}. This process will not in general preserve the separability of Alice and Charlie's measurements, but we do not need this below. Conversely, we do require separability of Bob's measurements which is why we leave his measurements unchanged.

To recap, we have shown that the existence of a behavior in $QB_2$ satisfying conditions \eqref{e:cond1}-\eqref{e:cond2} implies the existence of a (possibly different) behavior satisfying these conditions where the measured state is a pure state of form \eqref{e:state}, Alice and Charlie's measurements are projective, and Bob's measurements are separable as in Expression (1) in the main text. We now show that for a behavior induced this way, satisfaction of \eqref{e:cond1} is in fact incompatible with satisfaction of \eqref{e:cond2}.

\medskip

\noindent \textbf{Step 2: Implications of self-testing.} Bob's POVM on measurement setting $Y=1$ is given by $\{E_0,E_1\}$  where $E_0$ is of separable form
\begin{equation}\label{e:sepform}
E_0= \sum_{i=1}^m E^i_0 = \sum_{i=1}^m c_i\ket{\varphi_i}\bra{\varphi_i}\otimes\ket{\varphi'_i}\bra{\varphi'_i}
\end{equation}
where $\ket{\varphi_i}\bra{\varphi_i}$ acts on the state shared with Alice and $\ket{\varphi'_i}\bra{\varphi'_i}$ acts on the state shared with Charlie. Because of the tensor product structure of the measurements among the parties (or, relatedly, the no signaling principle), we can consider Bob to perform his measurement on \eqref{e:state} first, followed by Alice and Charlie measuring their resulting post-measurement state, which will be 
\begin{equation}\label{e:postbob}
\text{Tr}_B\left[\left(I_{A^{(b)}} \otimes E_0 \otimes I_{C^{(b)}} \otimes I_{AC}\right)\ket\psi\bra\psi\right]/\text{Prob}(E_0)
\end{equation}
where $\text{Prob}(E_0) = \text{Tr}\left[\left(I_{A^{(b)}} \otimes E_0 \otimes I_{C^{(b)}} \otimes I_{AC}\right)\ket\psi\bra\psi\right]$ and the notation $M_{P^{(q)}}$ indicates an operator on a register possessed by party $P$ that is (potentially) entangled with party $Q$. Note the expression above for the reduced state given in terms of the POVM element $E_0$, which is used as Equation (1) in \cite{bancal:2018}, is equivalent to (2.160) of \cite{NC} by the partial cyclicity of the partial trace \eqref{e:ptcyclic}.

Let us compute Equation~\eqref{e:postbob} explicitly. First we expand $\ket{\psi}_{AB} = \sum_l \lambda_l^{AB} \ket{\xi_l}_{A^{(b)}} \otimes \ket{\eta_l}_{B^{(a)}}$ and $\ket{\psi}_{BC} = \sum_{l^{\prime}} \lambda_{l^{\prime}}^{BC} \ket{\xi_{l^{\prime}}}_{B^{(c)}} \otimes \ket{\eta_{l^{\prime}}}_{C^{(b)}}$ in their Schmidt decompositions. Then the above (unnormalized) post-Bob's-measurement reduced state is given by 
\begin{align*}
\text{Tr}_B\left[\left(I_{A^{(b)}} \otimes E_0 \otimes I_{C^{(b)}} \otimes I_{AC}\right)\ket\psi\bra\psi\right] 
&=
\sum_i c_i\text{Tr}_B\left[\left(I_{A^{(b)}} \otimes \ket{\varphi_i}\bra{\varphi_i}_{B^{(b)}}\otimes\ket{\varphi'_i}\bra{\varphi'_i}_{B^{(c)}} \otimes I_{C^{(b)}} \otimes I_{AC}\right)\ket\psi\bra\psi\right] \\
&= \sum_{i=1}^m c_i \ket{x_i^{\prime}}\bra{x_i^{\prime}} \otimes\ket{y_i^{\prime}}\bra{y_i^{\prime}} \otimes \ket{\psi}\bra{\psi}_{AC},
\end{align*} where \begin{align*}
\ket{x_i^{\prime}} = \sum_l \lambda_l^{AB} \langle  \varphi_i | \eta_l \rangle \ket{\xi_l}_{A^{(b)}} \quad \text{ and } \quad
\ket{y_i^{\prime}} = \sum_{l^{\prime}} \lambda_{l^{\prime}}^{BC} \langle \varphi^{\prime}_{l^{\prime}}| \xi_{l^{\prime}} \rangle \ket{\eta_{l^{\prime}}}_{C^{(b)}}.
\end{align*}
Letting $
\ket{x_i} = \ket{x_i^{\prime}}/\|\ket{x_i^{\prime}}\|$ and $\ket{y_i} = \ket{y_i^{\prime}}/\|\ket{y_i^{\prime}}\|$ we get that the reduced Alice-Charlie-state, after Bob's measurement $Y=1$ and outcome $B=0$, is \begin{align}\label{e:postbob1}
\sum_i c_i^{\prime} \ket{x_i}\bra{x_i}_{A^{(b)}} \otimes \ket{y_i}\bra{y_i}_{C^{(b)}} \otimes \ket{\psi}\bra{\psi}_{AC},
\end{align} where $c_i^{\prime}$ are now modified positive scalars which sum to 1. Expression \eqref{e:postbob1} is thus equivalent to a convex mixture of $i$-indexed states. So whatever Alice and Charlie's measurements on \eqref{e:postbob1} are, these same measurements must produce a CHSH-maximizing behavior when applied to any of the individual $i$-indexed states appearing in \eqref{e:postbob1}, recalling that if an average of quantum-achievable behaviors maximally violates CHSH, each individual behavior must as well. Continuing the analysis for an individual state is simplified because each $i$-th state in \eqref{e:postbob1} is pure.

Now we are well-positioned to apply the self-testing argument to show that $\ket\psi_{AC}$ is effectively a Bell state. Fix a choice of $i$ in \eqref{e:postbob1}. Re-ordering terms, relabeling $\ket{x_i}_{A^{(b)}}$ and $\ket{y_i}_{C^{(b)}}$ as $\ket{0}_{A^{(b)}}$ and $\ket{0}_{C^{(b)}}$, and employing a Schmidt decomposition for $\ket\psi_{AC}$, Alice and Charlie's state is \begin{equation*}
    \ket{0}_{A^{(b)}}\otimes \left(\sum_j \gamma_j \ket{\psi_j}_{A^{(c)}}\otimes \ket{\psi_j}_{C^{(a)}}\right)\otimes\ket{0}_{C^{(b)}}.
\end{equation*} The self-testing construction of Figure 4 of \v{S}upi\'{c} and Bowles \cite{supic:2020} tells us that, since Alice and Charlie's measurements of this state maximally violates CHSH, then given the state 
\begin{align}
&\ket{0}_{a'}\otimes\ket{0}_{A^{(b)}}\otimes \left(\sum_j \gamma_j \ket{\psi_j}_{A^{(c)}}\otimes \ket{\psi_j}_{C^{(a)}}\right)\otimes\ket{0}_{C^{(b)}}\otimes\ket{0}_{c'}\notag\\
&= \sum_j \gamma_j \ket{0}_{a'}\otimes\ket{0}_{A^{(b)}}\otimes \ket{\psi_j}_{A^{(c)}}\otimes \ket{\psi_j}_{C^{(a)}}\otimes\ket{0}_{C^{(b)}}\otimes\ket{0}_{c'},\label{e:oldstate}
\end{align} where the adjoined $\ket 0$ states with the primed subscripts are qubits (the original $\ket 0$ states could be in higher dimensional spaces), there exist local unitaries $U$ and $V$ operating respectively on the first and last three registers such that $U\otimes V$ applied to the above state yields
\begin{equation}\label{e:firstxi}
\sum_{i\in\{0,1\}} \frac{1}{\sqrt 2} \ket i_{a'} \otimes \ket \xi \otimes \ket i_{c'} ;
\end{equation} that is, the Bell state $\ket {\Phi^+}=(\ket{00}+\ket{11})/\sqrt 2$ on the outer introduced qubits, tensored with a pure state $\ket \xi$ on the middle-four registers. To consider the constraints that this condition imposes on the form of the original state in \eqref{e:oldstate}, let us write $\ket \xi$ in a Schmidt decomposition of $\sum_{k=1}^{n} \delta_k \ket {A_k} \ket {C_k}$; in general it is possible that the vectors $\ket {A_k}$ are entangled over Alice's two subsystems, and similarly for $\ket {C_k}$. With this we re-write \eqref{e:firstxi} as 
\begin{equation}\label{e:targets}
\sum_{i\in\{0,1\}}\sum_{k=1}^{n} \frac{\delta_k }{\sqrt 2} \ket i \otimes  \ket {A_k} \otimes \ket {C_k} \otimes \ket i
\end{equation}
which considered as a single sum of $2n$ terms consists of real positive coefficients $\delta_k/\sqrt 2$ of orthogonal sets $\{\ket i \otimes  \ket {A_k}\}_{i,k}$ and $\{  \ket {C_k}\otimes \ket i\}_{i,k}$, and so is itself a Schmidt decomposition. Now, let us consider what happens when we apply the inverse map $U^\dagger \otimes V^\dagger$ to this state. The result will be a state of the form 
\begin{equation}\label{e:latestate}
\sum_{j=1}^{2n}\lambda_j\ket {A'_j} \otimes \ket {C'_j},
\end{equation}
again a Schmidt decomposition as the $\ket {A'_j}$ and $\ket {C'_j}$ are orthogonal due to the unitarity of $U^\dagger$ and $V^\dagger$. Now because \eqref{e:latestate} is the same state as \eqref{e:oldstate}, each state $\ket {A'_j}$ must be of the form $\ket 0 \ket 0 \ket \varphi$ for some state $\ket \varphi$. (One way to see this is to observe that the partial trace $\rho_C$ of the state in \eqref{e:latestate} is $\sum_j |\lambda_j|^2\ket {A'_j}\bra {A'_j}$ which is a diagonal representation that must be equal to the diagonal representation $\rho_A=\sum_j |\gamma_j|^2 \ket{0}_{a'}\ket{0}_{A^{(b)}} \ket{\psi_j}_{A^{(c)}}\bra{0}_{a'}\bra{0}_{A^{(b)}} \bra{\psi_j}_{A^{(c)}} $ obtained from \eqref{e:oldstate}; the second representation demonstrates that all non-null eigenspaces of $\rho_A$ are spanned by vectors of the form $\ket 0 \ket 0 \ket \varphi$, and since the $\ket {A'_j}$ belong to these eigenspaces, they must be of this form as well.) Thus re-writing each $\ket {A'_j}$ as $\ket 0 \ket 0 \ket {\psi^a_j}$, where we remark the $\ket {\psi^a_j}$ must be orthogonal, and doing similarly for the $\ket {C'_j}$, we see that the state \eqref{e:oldstate} admits a (possibly modified/reordered) Schmidt decomposition of the form 
\begin{equation}\label{e:needtoo}
\sum_{j=1}^{2n} \gamma_j \ket{0}_{a'}\otimes\ket{0}_{A^{(b)}}\otimes \ket{\psi^a_j}_{A^{(c)}}\otimes \ket{\psi^c_j}_{C^{(a)}}\otimes\ket{0}_{C^{(b)}}\otimes\ket{0}_{c'}
\end{equation} such that each term in the above summand maps through $U\otimes V$ to a distinct term in the summand \eqref{e:targets}, and importantly we observe that half of these -- assume, without loss of generality, those with indices $j\in\{1,...,n\}$ -- are mapping to terms with $\ket 0$s in the $a'$/$c'$ registers, while the remaining half maps to the $\ket 1$ terms. Combining terms within the two groups together as $\ket{\psi_0}=\sqrt 2\sum_{j=1}^n \gamma_j\ket{\psi^a_j}\otimes\ket{\psi^c_j}$ and $\ket{\psi_1}=\sqrt 2\sum_{j=n+1}^{2n} \gamma_j\ket{\psi^a_j}\otimes\ket{\psi^c_j}$, we can re-write the state in \eqref{e:needtoo} as 
\begin{equation}\label{e:Fdef}
\ket F = \sum_{k\in\{0, 1\}}\frac{1}{\sqrt 2} \ket{0}_{a'}\otimes\ket{0}_{A^{(b)}}\otimes  \ket{\psi_k}_{AC}\otimes\ket{0}_{C^{(b)}}\otimes\ket{0}_{c'}
\end{equation}
with the two summands mapping to $\frac{1}{\sqrt 2}\ket 0 \otimes \ket \xi \otimes \ket 0$ and $\frac{1}{\sqrt 2}\ket 1 \otimes \ket \xi \otimes \ket 1$, respectively, under the map $U\otimes V$. We note the appearance of the prefactor of $1/\sqrt 2$ corresponds the states $\ket{\psi_k}_{AC}$ being normalized, as follows from the length-preserving property of $U\otimes V$ and the fact that $\ket 0 \otimes\ket \xi\otimes \ket 0$ and $\ket 1 \otimes\ket \xi\otimes \ket 1$ are unit length. 

Equation \eqref{e:Fdef} captures the precise manner in which Alice and Bob's shared state $\ket \psi_{AC}$ has the essence of a Bell state. We continue the self-testing analysis to formulate how Alice's measurement effectively ignores the $A^{(b)}$ register shared with Bob, acting only on $\ket\psi_{AC}$. Let us denote Alice's measurement on setting $X=0$ with the projectors $\{\Pi_0,\Pi_1\}$ corresponding to outcomes $0$ and $1$. By Equation (39) of \v{S}upi\'{c} and Bowles \cite{supic:2020} (the roles of Alice and Bob are exchanged here), we can say that $\Pi_0$ is the $\sigma^+_z$ operator in the following sense, where $I_C= I_{C^{(a)}}\otimes I_{C^{(b)}}$:
\begin{align*}
(U\otimes V)(I_{a'}\otimes \Pi_0 \otimes I_{C} \otimes I_{c'}) \ket F = (\ket 0 \bra 0 \otimes I_{\ket \xi} \otimes I_{c'}) \sum_{i\in\{0,1\}}\frac{1}{\sqrt 2} \ket i \otimes \ket \xi \otimes \ket i =\frac{1}{\sqrt 2} \ket 0 \otimes \ket \xi \otimes \ket 0
\end{align*} and so applying $U^\dagger\otimes V^\dagger$ to both sides we see \begin{equation*}
    (I_{a'}\otimes\Pi_0 \otimes I_C\otimes I_{c'}) \ket F = \ket{0}_{a'}\otimes\ket{0}_{A^{(b)}}\otimes  \ket{\psi_0}_{AC}\otimes\ket{0}_{C^{(b)}}\otimes\ket{0}_{c'},
\end{equation*} which implies, along with a parallel argument for $\Pi_1$, that (now disregarding the introduced states $\ket 0_{a'}$ and $\ket 0_{b'}$) we have
\begin{equation}\label{e:splitform}
\Pi_i \otimes I_C \left(\sum_{k\in\{0,1\}}\frac{1}{\sqrt 2}\ket{0}_{A^{(b)}}\otimes  \ket{\psi_k}_{AC}\otimes\ket{0}_{C^{(b)}}\right)=\frac{1}{\sqrt 2}\ket{0}_{A^{(b)}}\otimes  \ket{\psi_i}_{AC}\otimes\ket{0}_{C^{(b)}}
\end{equation} for both choices of $i \in \{0,1\}$. With Alice's measurement ``ignoring'' $\ket 0_{A^{(b)}}$ in this way, while yielding 50-50 coin toss via a measurement on the portion shared with Charlie, it would impossible for Bob to guess Alice's outcome with perfect probability. 

The complete picture is, however, more complicated than \eqref{e:splitform}: first, carrying through the above analysis with a different choice of $i$ in \eqref{e:postbob1} associated with Bob's measurement outcome $B=0$ on setting $Y=1$ could lead to a different (though analogous) form of \eqref{e:splitform}: the state $\ket 0 _{A^{(b)}}$ could be different, and furthermore $\ket\psi_{AC}$ could ``split" into two different halves $\ket{\psi^*_0}_{AC}$ and $\ket{\psi^*_1}_{AC}$. Second, while Alice and Charlie's post-Bob-measurement state in \eqref{e:postbob1} will be a convex mixture of such analogous states, their pre-Bob-measurement state will be something else, and it is that pre-measurement state that Bob will be confronted with when trying to devise a measurement on setting $Y=0$ to align with Alice's.

\medskip

\noindent \textbf{Step 3: Characterizing Alice's $A^{(b)}$ register.} To address these issues, we will show that Alice's $A^{(b)}$
register in \eqref{e:splitform} admits an orthogonal decomposition into subspaces 
\begin{equation}\label{e:orthog}
A^{\mathrm{split}}\oplus A^{\mathrm{rest}} = A^{\mathrm{split}}_0 \oplus\cdots \oplus A^{\mathrm{split}}_K \oplus A^{\mathrm{rest}}
\end{equation}
such that $\Pi_i \otimes I_{C^{(a)}}$ ``splits'' $\ket a_{A^{(b)}}\otimes\ket \psi_{AC}$ as in \eqref{e:splitform} whenever $\ket a_{A^{(b)}}$ lies in $A^{\mathrm{split}}_0$, whereas if $\ket a_{A^{(b)}}$ lies in a different $A^{\mathrm{split}}_j$ the state splits as in \eqref{e:splitform} but with a different splitting of $\ket\psi_{AC}$ into distinct halves $\ket {\psi^*_k}_{AC}$, one $A^{\mathrm{split}}_j$ for each potential splitting.  The action of $\Pi_i$ for $\ket a_{A^{(b)}}$ lying in $A^{\mathrm{rest}}$ is uncharacterized but we can show Alice's pre-Bob-measurement state must contain nonzero amplitudes in the $A^{\mathrm{split}}$ component, making condition \eqref{e:cond2} impossible. We remark that the possibility of multiple distinct $A_j^{\mathrm{split}}$ spaces cannot be discounted as it can occur if Alice and Charlie share multiple separate singlets jointly comprising $\ket \psi_{AC}$ and don't always measure the same one; we provide an explicit example in fuller detail after the conclusion of the proof.

To arrive at \eqref{e:orthog}, let us first consider the collection of first-register states that lead to the same splitting of $\ket \psi_{AC}$ as in \eqref{e:splitform}; i.e., define $A_0^{\mathrm{split}}$ as the subset of states $\ket  a_{A^{(b)}}$ for which 
\begin{equation*}
\Pi_i \otimes I_{C^{(b)}} \left(\sum_{k\in\{0,1\}}\frac{1}{\sqrt 2}\ket{a}_{A^{(b)}}\otimes  \ket{\psi_k}_{AC}\right)=\frac{1}{\sqrt 2}\ket{a}_{A^{(b)}}\otimes  \ket{\psi_i}_{AC}.
\end{equation*}
(Note that we are safely ignoring the extra register $\ket 0 _{C^{(b)}}$ that appears in \eqref{e:splitform}; a state satisfies the above condition if and only if it satisfies the same condition with the extra register included.) It is straightforward to check that $A^{\mathrm{split}}_0$ is closed under linear combinations and is thus a sub\textit{space} of dimension $k\ge 1$. For each alternate possible splitting into different halves $\ket {\psi^*_k}_{AC}$, we can define a different $A_j^{\mathrm{split}}$, which must also be a subspace. What is not apparent \textit{a priori} is that these different subspaces must be orthogonal as claimed in \eqref{e:orthog}.

To prove that the various $A_j^{\mathrm{split}}$ are orthogonal, we use the following refined observation about the action of $\Pi_i$ in \eqref{e:splitform}. Recalling that the $\ket{\psi_k}$ in \eqref{e:splitform} can be expressed as sums of states of the form $\ket{\psi^a_j}\otimes \ket{\psi^c_j}$ as in \eqref{e:needtoo}, we observe that $\Pi_0$ must preserve these individual Alice states $\ket{0}_{A^{(b)}}\otimes \ket{\psi^a_j}_{A^{(c)}}$ for $j\in\{1,...,n\}$ while annihilating the $j\in\{n+1,...,2n\}$ states, and vice versa for $\Pi_1$. To see why this is, in \eqref{e:splitform} expand all the the $\ket{\psi_k}$ in terms of $\ket{\psi^a_j}\otimes\ket{\psi^c_j}$ and apply the linearity of $\Pi_0$ on the left side to obtain the equality \begin{equation*}
\sum_{j=1}^{2n} \gamma_j \left[\Pi_0 \left(\ket{0}_{A^{(b)}}\otimes \ket{\psi^a_j}_{A^{(c)}}\right)\right] \otimes \ket{\psi^c_j}_{C^{(a)}}\otimes\ket{0}_{C^{(b)}}=\sum_{j=1}^{n} \gamma_j \ket{0}_{A^{(b)}}\otimes \ket{\psi^a_j}_{A^{(c)}} \otimes \ket{\psi^c_j}_{C^{(a)}}\otimes\ket{0}_{C^{(b)}}.
\end{equation*} Then applying the projector $I_A\otimes \ket{\psi^c_{j^{\prime}}}\bra{\psi^c_{j^{\prime}}}_{C^{(a)}} \otimes \ket{0}\bra{0}_{C^{(b)}}$ on both sides of the above equation for each fixed ${j^{\prime}}$ yields the desired result, recalling that the different $\ket{\psi^c_j}$ are orthogonal as they arise from a Schmidt decomposition.

The above observation is useful because it allows us to show that $\Pi_0$ maps any product state with a $A^{(b)}$ register lying in the orthogonal complement of $A^{\mathrm{split}}_0$ to a vector whose $A^{(b)}$ components remain in the orthogonal complement of $A^{\mathrm{split}}_0$: let $\ket 0_{A^{(b)}}, ... , \ket {k-1}_{A^{(b)}}$ be an orthonormal basis of $A^{\mathrm{split}}_0$ and extend this to a complete orthonormal basis  with vectors $\ket i_{A^{(b)}}$, $i\ge k$. Consider the expansion of $\Pi_0$ as a sum of ket-bras in the orthonormal product basis of all states of the form $\ket i_{A^{(b)}} \otimes \ket {\psi^a_j}_{A^{(c)}}$, $j \in \{1,..., 2n\}$. (We safely ignore possible additional dimensions of the $AC$ state space, since the state $\ket \psi_{AC}$ does not have any components in those dimensions and so they are irrelevant.) This sum form of $\Pi_0$ will include the terms $\ket i \ket {\psi^a_j}\bra i  \bra {\psi^a_j}$ with $i$ ranging from $0$ to $k-1$ and $j$ ranging from $1$ to $n$, while no other additional terms can have the form $c\ket x \ket{\psi^a_y} \bra i \bra {\psi^a_j}$ for $i \in \{0,...,k-1\}$, which would contradict $\Pi_0$'s action on $\ket i  \ket {\psi^a_j}$ states as discussed in the previous paragraph.  By the self-adjointness of $\Pi_0$, this importantly rules out terms of the form $c\ket i \ket {\psi^a_j}\bra x \bra {\psi^a_y}$ for $i \in \{0,...,k-1\}$  and $x \ge k$. Hence $\Pi_0$ maps any state having a first register lying in the orthogonal complement of $A^{\mathrm{split}}_0$ to a vector whose first register components remain in the orthogonal complement of $A^{\mathrm{split}}_0$. Naturally, this argument will hold for $\Pi_1$ and any other $A^{\mathrm{split}}_0$.

We can now show that if $\ket a \in A^{\mathrm{split}}_j$ for $j\ne 0$, then $\ket a \perp A^{\mathrm{split}}_0$. We prove this claim as follows: it is always possible to express $\ket a$ in the form $\alpha \ket a ^{N} + \beta \ket a^{N^\perp}$ with $\ket a ^{N} \in A^{\mathrm{split}}_0$ and $\ket a^{N^\perp}\perp A^{\mathrm{split}}_0$; the claim holds if this expression requires $ \ket a ^N = \vec 0$ or $\alpha = 0$. We can write
\begin{align*}
(1/\sqrt 2)(\alpha \ket a ^{N} + \beta \ket a^{N^\perp} ) \otimes \ket {\psi_0^*}_{AC}
&=(1/\sqrt 2)\ket a \otimes \ket {\psi_0^*}_{AC}\\
&=\left(\Pi_0 \otimes I_{C^{(a)}} \right)\ket a \otimes \ket \psi_{AC} \\
&= \left(\Pi_0 \otimes I_{C^{(a)}} \right) \alpha \ket a ^{N} \otimes \ket \psi_{AC}+\left(\Pi_0 \otimes I_{C^{(a)}} \right) \beta \ket a^{N^\perp}\otimes \ket \psi_{AC}\\
&= \frac{\alpha}{\sqrt 2} \ket a ^{N}\otimes\ket {\psi_0}_{AC} + \beta \left(\Pi_0 \otimes I_{C^{(a)}}\right) \ket  a^{N^\perp}\otimes \ket \psi_{AC}.
\end{align*}
Applying the projector $(P^{N^\perp})_{A^{(b)}}\otimes I_{AC}$, where $P^{N^\perp}$ is the projector onto $(A^{\mathrm{split}}_{0})^\perp$, to both sides above yields
\begin{equation}\label{e:necess}
\frac{\beta}{\sqrt 2} \ket  a^{N^\perp}  \otimes \ket {\psi_0^*}_{AC} =\beta \left(\Pi_0 \otimes I_{C^{(a)}} \right) \ket a^{N^\perp}\otimes \ket \psi_{AC},
\end{equation}
using the fact that $\Pi_0\otimes I_{C^{(a)}}$ maps states with first register in $(A^{\mathrm{split}}_{0})^\perp$ to states with first register in $(A^{\mathrm{split}}_{0})^\perp$, on which $P^{N^\perp}\otimes I_{AC}$ will act as identity. Now substituting \eqref{e:necess} into the preceding equality requires $|\alpha| = 0$ or $\ket \varphi ^{N}= \vec 0$, recalling that $\ket {\psi_0^*}_{AC} \ne \ket {\psi_0}_{AC}$.

We now have the orthogonal decomposition of \eqref{e:orthog}: all $A_j^{\mathrm{split}}$ subspaces must be orthogonal by the arguments above, and we can take $A^{\mathrm{rest}}$ be the orthogonal complement of the union of all such subspaces. By construction $A^{\mathrm{rest}}$ does not have the splitting property, so conditioned on Bob seeing outcome $B=0$ given setting $Y=1$, Alice's first register in the post-measurement state is contained in $A^{\mathrm{split}}=A^{\mathrm{split}}_0\oplus \cdots \oplus A^{\mathrm{split}}_K$.

\medskip

\noindent \textbf{Step 4: Obtaining a bound on Bob's alignment probability.} Alice's state \textit{before} Bob performs measurement $Y=1$ and observes outcome $B=0$ could have positive amplitudes on states with $A^{(b)}$ register outside $A^{\mathrm{split}}$; for such states, Alice's measurement may not act trivially on her portion with Bob and/or it might measure $\ket{\psi}_{AC}$ differently. This means that when Bob chooses to measure $Y=0$, he may have nontrivial opportunities to align his measurement outcome with Alice. However, we can demonstrate that the sum of the magnitude of Alice's amplitudes on $A^{\mathrm{split}}$-type states must be bounded below by $P(B=0|Y=1)>0$, which is sufficient to show Bob cannot align his $Y=0$ outcomes with Alice perfectly. To proceed, expand the Alice-Bob state $\ket \psi_{AB}$ in a product basis such that Alice's basis aligns with the orthogonal subspaces of \eqref{e:orthog}, permitting an expression
\begin{equation}\label{e:orthexp}
\ket{\psi^{AB}}
= \sum_{j \in A^\mathrm{split}}\gamma_j \ket{a_j}_{A^{(b)}} \ket{b_j}_{B^{(a)}}+\sum_{j\in A^\mathrm{rest}}\gamma_j\ket{a_j}_{A^{(b)}} \ket {b_j}_{B^{(a)}}
\end{equation}
where the $\ket{a_j} $ are orthogonal though the $\ket{b_j}$ are not necessarily, and now, recalling the representation of Bob's separable POVM element from \eqref{e:sepform}, we can write
\begin{align}
 P(B=0|Y=1)
&=\text{Tr}\left[I_{A^{(b)}} \otimes  E_0 \otimes I_{C^{(b)}} \otimes I_{AC} \ket \psi \bra \psi\right]\notag\\
&=\sum_i\text{Tr}\left[I_{A^{(b)}} \otimes  E^i_0 \otimes I_{C^{(b)}} \otimes I_{AC} \ket \psi \bra \psi\right]\notag\\
&=\sum_i\text{Tr}\left[\left( I_{A^{(b)}} \otimes  \sqrt{E_0^i}\otimes I_{C^{(b)}} \otimes I_{AC} \right ) \ket \psi \bra \psi \left (I_{A^{(b)}} \otimes  \sqrt{E_0^i}\otimes I_{C^{(b)}} \otimes I_{AC}\right)\right]\label{e:bigtrace}
\end{align}
using \eqref{e:ptcyclic} in the last line where the square root of $E_0^i$ given by the simple expression $\sqrt{c_i} \ket{\varphi_i}\bra{\varphi_i}\otimes \ket{\varphi'_i}\bra{\varphi'_i}$. Now we can simplify \eqref{e:bigtrace} by writing out $\ket\psi$ as in \eqref{e:state} with $\ket\psi_{AB}$ expanded as in \eqref{e:orthexp} and observing that each $I_{A^{(b)}}\otimes \sqrt{c_i} \ket{\varphi_i}\bra{\varphi_i}_{B^{(a)}}\otimes \ket{\varphi'_i}\bra{\varphi'_i}_{B^{(c)}}\otimes I_{C^{(b)}}\otimes I_{AC}$ must map all terms of the form $\gamma_j \ket{a_j}_{A^{(b)}} \ket {b_j}_{B^{(a)}} \ket \psi_{BC}  \ket \psi_{AC}$ for ${j \in A^\mathrm{rest}}$ to $\vec 0$, since otherwise Alice's post measurement state would contain $A^\mathrm{rest}$ components in the first register, and such states are incompatible with a maximal CHSH violation. Thus we can replace $\ket \psi$ in \eqref{e:bigtrace} with \begin{equation*}
    \ket {\psi'}=\left(\sum_{j \in A^\mathrm{split}}\gamma_j \ket{a_j}_{A^{(b)}} \ket{b_j}_{B^{(a)}}\right)\otimes \ket \psi_{BC} \otimes \ket \psi_{AC}
\end{equation*} to get an equivalent expression; applying \eqref{e:ptcyclic} in reverse and pulling the sum back into the expression yields
\begin{align}
P(B=0|Y=1)
&=\text{Tr}\left[I_{A^{(b)}} \otimes  E_0 \otimes I_{C^{(b)}} \otimes I_{AC} \ket {\psi'} \bra {\psi'}\right]\notag\\
&=||\ket {\psi'}||^2\text{Tr}\left[I_{A^{(b)}} \otimes  E_0 \otimes I_{C^{(b)}} \otimes I_{AC} \left(\frac{\ket{\psi'}}{||\ket {\psi'}||} \right)\left(\frac{\bra {\psi'}}{||\ket {\psi'}||}\right)\right]\notag\\
&\le ||\ket {\psi'}||^2\notag\\
&=\sum_{j \in A^\mathrm{split}}|\gamma_j|^2\label{e:ampbound}
\end{align}
where the inequality follows from the fact that the trace expression is a probability, corresponding to a measurement of the quantum state given by the normalized form of $\ket {\psi'}$. This is our desired lower bound on the amplitudes on the $A^\mathrm{split}$ states.

We now finish the argument by showing that, considering Alice to perform measurement $X=0$ first, there is a non-trivial overlap in Bob's two potential reduced states (corresponding to Alice's two different outcomes) such that he cannot distinguish between her outcomes perfectly. Utilizing \eqref{e:orthexp} to write the pre-measurement state \begin{equation*}
\left(\sum_{j \in A^\mathrm{split}}\gamma_j \ket{a_j} \ket {b_j}\right)\otimes \ket \psi_{BC} \otimes \ket \psi_{AC}+ \left(\sum_{j \in A^\mathrm{rest}}\gamma_j \ket{a_j} \ket {b_j}\right)\otimes \ket \psi_{BC} \otimes \ket \psi_{AC},
\end{equation*} the result of Alice's measurement $\{\Pi_i\}_{i\in \{0,1\}}$ given outcome $i$ is a subnormalized state that can be written as \begin{equation}\label{e:postsubnorm}
\sum_{j\in A^\mathrm{split}}\gamma_j \ket{a_j}_{A^{(b)}} \ket {b_j}_{B^{(a)}}\otimes \ket \psi_{BC} \otimes \left ( \frac{1}{\sqrt 2}\ket{\psi^j_i}_{AC}\right) + \delta^\mathrm{rest}_i \ket{\psi^i}^\mathrm{rest}_{ABC} 
\end{equation} where we do not know too much about the form of the normalized state $\ket{\psi}^\mathrm{rest}_{ABC}$, though we do know that it can have $\ket{a_j}_{A^{(b)}}$ components only in $A^\mathrm{rest}$ when expanded with this basis, and so $\ket{\psi}^\mathrm{rest}_{ABC}$ is orthogonal to all the vectors in the first sum which are in turn orthogonal to each other. This orthogonality allows us to say that the outcome $i$ occurs with probability $P(i)=\sum_{j\in A^\mathrm{split}}|\gamma_j|^2/2 + |\delta^\mathrm{rest}_i|^2$, with only the second term depending on $i$. Now tracing out \eqref{e:postsubnorm} over $A$ and $C$ to obtain Bob's reduced state, and again taking advantage of the orthogonality of each summed term in \eqref{e:postsubnorm}, we see that Bob's normalized reduced state given Alice's outcome $i$ is \begin{equation*}
\rho_B=\frac{1}{P(i)}\left[\left(\sum_{j\in A^\mathrm{split}}\frac{|\gamma_j|^2}{2} \ket{b_j}\bra{b_j} \otimes \mathrm{tr}_C(\ket \psi_{BC}\bra\psi_{BC}) \right)+ |\delta_i^\mathrm{rest}|^2\mathrm{tr}_{AC}(\ket{\psi^i}^\mathrm{rest}_{ABC}\bra{\psi^i}^\mathrm{rest}_{ABC})\right].
\end{equation*} This reduced state is equivalent to a convex combination $p_i\rho+q_i\tau_i$ of two density operators where $\rho$ is independent of $i$ and $p_i= [P(i)]^{-1}\sum_{j\in A^\mathrm{split}}|\gamma_j|^2/2$, $q_i=1-p_i$. There is no measurement allowing Bob to perfectly distinguish between these two reduced states, and we have, using the shorthand $N=\sum_{j\in A^\mathrm{split}}|\gamma_j|^2$ and $T_i = |\delta_i^\mathrm{rest}|^2$,
\begin{align*}
P(A\ne B|X=0,Y=0) &= P(A=0,B=1|X=0,Y=0)+P(A=1,B=0|X=0,Y=0)\\
&=\sum_{i\in\{0,1\}}[p_iP(B\ne i|\rho)+q_iP(B\ne i|\tau_i)]P(A=i|X=0)\\
&\ge [p_0P(B=1|\rho)](N/2+T_0)+[p_1P(B=0|\rho)](N/2+T_1)\\
&=(N/2)P(B=1|\rho)+(N/2)P(B=0|\rho)\\
&=N/2\\
&\ge P(B=0|Y=1)/2,
\end{align*}
with the last inequality following from \eqref{e:ampbound}.\hfill $\Box$

\medskip

In the proof, we remarked after Eq.~\eqref{e:orthog} that the possibility of multiple distinct $A_j^{\mathrm{split}}$ spaces comprising $A^{\mathrm{split}}$ cannot be discounted. To aid intuition, we provide an example of a scenario in which this will occur. Let Alice and Charlie share two separate singlets jointly comprising $\ket \psi_{AC}$. Alice performs CHSH measurements on her portion of the first $\ket \psi_{AC}$ singlet if a measurement of a (separate) qubit maximally entangled with Bob yields ``0", whereas she performs CHSH measurements on the second $\ket \psi_{AC}$ singlet if the Bob-linked measurement yields ``1". Charlie employs a parallel strategy, also using a qubit maximally entangled with Bob to govern which of the $\ket \psi_{AC}$ singlets he chooses to measure. Bob thus possesses a qubit entangled with Alice and 
a qubit entangled with Charlie. If Bob measures these in the computational basis and sees both as 0, or both as 1, he knows Alice and Charlie are measuring the same singlet, upon which he reports outcome $B=0$ (which will occur 50\% percent of the time) leading to a maximal Alice-Charlie CHSH violation conditioned on this outcome. Depending on which singlet is being measured, the \eqref{e:splitform} ``split" of the single state $\ket \psi_{AC}$, which comprises both singlets shared by Alice and Charlie, will not be the same: the $A^{(b)}$ and $C^{(b)}$ registers will lie in different $A_j^{\mathrm{split}}$ spaces, and the two halves $\ket {\psi_k}_{AC}$ that sum to $\ket \psi_{AC}$ will be different.

\subsection{Satisfying the conditions of Theorem 1 with a GHZ state and no entangled measurement}\label{app:Thm-1-with-GHZ}

We can induce a behavior $P(ABC|XYZ)$ satisfying the conditions of Theorem 1 without entangled measurements by distributing the three-qubit entangled GHZ state $(\ket{000}+ \ket{111})/\sqrt 2$ to all three players. Let Bob's measurement for setting $Y=1$ be the projectors onto $\ket + = (\ket 0 + \ket 1)/\sqrt 2$ and $\ket - = (\ket 0 - \ket 1)/\sqrt 2$. Then conditioned on Bob's observation of $\ket +$, which occurs with positive probability $P(B=0|Y=1)=1/2$, Alice and Charlie possess the Bell state $\ket{\Phi^+}=(\ket{00}+\ket{11})/\sqrt 2$. This can yield a maximal CHSH value while Alice measures $\sigma_z$ for setting $X=0$; i.e.,~she is measuring in the computational basis $\{\ket 0, \ket 1\}$. Finally, if Bob also measures in the computational basis when his setting is $Y=0$, the condition $P(A=B|X=0,Y=0)=1$ is met.

\subsection{Dilation of POVMs to projective measurements}\label{app:POVM-dilation}
In this section, we reproduce the argument in Section 9-6 of \cite{PeresQT} which shows that a POVM can be replaced with a projection valued measurement (PVM) without modifying the structure of the state. This construction is different from the standard construction, for instance Section 2.2.8 in \cite{NC}. In the following, $[K]$ denotes the set $\{1,..,K\}$ of the first $K$ positive integers.

\medskip

\textbf{Lemma 1}\label{lem:rank-1-POVM-to-PVM} Let $ H$ be a $d$-dimensional Hilbert space, and suppose $\{\ket{x_k}\}_{k\in [K]}$ (for some $K\in \mathbb N$ with $K > d$) are non-zero vectors such that $\sum_{k=1}^K \ket{x_k}\bra{x_k} = I_{ H}$. Let $r = K - d$. Then, for each $k\in [K]$, there exists a vector $\ket{\varphi_k} \in \mathbb C^r$, such that the set $\{\ket{y_k} := \ket{x_k} + \ket{\varphi_k} \}_{k\in [K]} \subseteq  H \oplus \mathbb C^r$ forms an orthonormal set.

\medskip

\textit{Proof}. Fix an orthonormal basis $\{\ket{l}\}_{l=0}^{d-1}$ of $ H$, and for each $k\in [K]$ let $\ket{x_k} = \sum_{l=0}^{d-1} x_{k,l}\ket{l}$, where $x_{k,l}\in \mathbb C$. Writing the equation $\sum_{k=1}^K \ket{x_k}\bra{x_k} = I_{ H}$ in component-form, we get \begin{equation}\label{eq:first-d-ons}
		\sum_{k=1}^K x_{k,i}\overline{x_{k,j}} = \delta_{i,j}, \qquad i,j\in \{0,..,d-1\}.
	\end{equation} For each $k\in [K]$, we want to show that there are choices of constants $\lambda_{k,l}$ for which $\ket{\varphi_k} := \sum_{l=d}^{K-1} \lambda_{k,l}\ket{l}$ satisfies the claim of the Lemma, where we are using $\ket d, ..., \ket {K-1}$ as a basis for $\mathbb C^r$. The set $\{\ket{y_k}\}_{k=1}^K$ is orthonormal if and only if $\langle  y_k, y_l\rangle = \delta_{k,l}$ for all $k,l\in [K]$, that is, \begin{equation}\label{eq:UU-star-is-id}
		\delta_{k,l} = \langle x_k| x_l \rangle  + \langle \varphi_k| \varphi_l \rangle = \sum_{i=0}^{d-1} \overline{x_{k,i}}x_{l,i} + \sum_{i=d}^{K-1}\overline{\lambda_{k,i}}\lambda_{l,i}, \qquad k,l\in [K].
	\end{equation}
	
	Consider the following $K\times K$ scalar matrix: \begin{equation*}
		U = \begin{bmatrix}
			x_{1,0} & x_{1,1} & \dots & x_{1,d-1} & \lambda_{1,d} & \lambda_{1,d+1} & \dots & \lambda_{1,K-1} \\
			x_{2,0} & x_{2,1} & \dots & x_{2,d-1} & \lambda_{2,d} & \lambda_{2,d+1} & \dots & \lambda_{2,K-1} \\
			\vdots & \vdots & \vdots & \vdots & \vdots & \vdots & \vdots & \vdots \\
			x_{K,0} & x_{K,1} & \dots & x_{K,d-1} & \lambda_{K,d} & \lambda_{K,d+1} & \dots & \lambda_{K,K-1} \\
		\end{bmatrix}.
	\end{equation*} Then, $U$ is a unitary matrix $\Leftrightarrow$ $UU^* = I_K$ $\Leftrightarrow$ Equation~\eqref{eq:UU-star-is-id} holds $\Leftrightarrow$ $\{\ket{y_k}\}_{k=1}^K$ is an orthonormal set. Moreover, Equation~\eqref{eq:first-d-ons} tells that the first $d$ columns of $U$ are orthonormal. It is then clear that the existence of vectors $\ket{\varphi_k}\in \mathbb C^r$ corresponds to extending the first $d$ orthonormal columns to an orthonormal basis of $\mathbb C^K$, which is always possible. \hfill $\Box$

\medskip

Above, $\ket{y_k}\bra{y_k}$ are rank-one projection operators on the larger space $\mathbb C^K$ whose actions on vectors wholly contained in the subspace $\mathbb C^d$ are identical to the actions of the $\ket{x_k}\bra{x_k}$, which themselves constitute a POVM with rank-one elements on the subspace. The following results show how to add the extra needed dimensions by introducing a tensored qudit, while also extending the result to POVMs with general elements. We use the concept of an \textit{isometry}: a linear map $V:H_1\to H_2$ between Hilbert spaces satisfying $\braket{\varphi|\psi}_{H_1} = \bra \varphi V^* V
\ket \psi_{H_2}$ for all choices of $\ket \varphi$ and $\ket \psi$ in $H_1$. In particular, the map $V:H \to H\otimes \mathbb C^r$ given by $V\ket \xi =\ket \xi \otimes \ket 0$, where $\ket 0$ is a fixed basis element of $\mathbb C^r$, is an isometry satisfying \begin{equation}\label{e:isomproperty}
V^*\left(\ket \xi \otimes \ket i \right)=\sum_j\ket j \bra j \big [V^* \left( \ket \xi \otimes \ket i\right)\big] =\sum_j \ket j \bra{Vj}\big ( \ket \xi \otimes \ket i\big) = \sum_j \ket j \braket{j|\xi}\braket{0|i}=\begin{cases} \ket \xi  & \text{ if } i =0\\ \mathbf{0} &\text { if } i \in \{1,...,r-1\}
\end{cases}
\end{equation}
for all $\ket \xi \in H$, where the $\ket j$ are an orthonormal basis of $H$. Consequently,
\begin{equation}\label{e:isomiden}
VV^* = I_H \otimes \ket 0 \bra 0 \qquad \text{and} \qquad V^*V = I_{H}
\end{equation}

\medskip

\textbf{Lemma 2}. Let $ H$ be a $d$-dimensional Hilbert space, and suppose $\{\ket{x_k}\}_{k\in [K]}$ (for some $K\in \mathbb N$ with $K > d$) are non-zero vectors such that $\sum_{k=1}^K \ket{x_k}\bra{x_k} = I_{ H}$. Let $r = K - d + 1$ and fix a unit vector $\ket\eta\in  H$. Then, for each $k\in [K]$, there exists a vector $\ket{\varphi_k} \in \mathbb C^r$ orthogonal to the basis vector $\ket 0\in \mathbb C^r$, such that the set \begin{equation}\label{eq:rank-1-PVM-set}
		 Y = \{\ket{y_k} := \ket{x_k} \otimes \ket{0} + \ket\eta \otimes \ket{\varphi_k} \}_{k\in [K]} \subseteq  H \otimes \mathbb C^r
	\end{equation} forms an orthonormal set. Moreover, if $V: H \to  H \otimes \mathbb C^r$ is the isometry defined by $V\ket{\xi} = \ket{\xi} \otimes \ket{0}$ (for all $\ket{\xi}\in  H$), then one has $V^*\ket{y_k} = \ket{x_k}$ for all $k\in [K]$. 

\medskip

\textit{Proof}. Applying Lemma~1, we get vectors $\ket{\widetilde{\varphi}_k}\in \mathbb C^{K-d}$ (for $k\in [K]$) such that the set $\{\ket{\widetilde{y}_k}:= \ket{x_k} + \ket{\widetilde{\varphi}_k}\}_{k\in [K]}$ is orthonormal in $ H \oplus \mathbb C^{K-d}$. For each $k\in [K]$, let $\ket{\widetilde{\varphi}_k} = \sum_{i=0}^{K-d-1} \lambda_{k,i}\ket{i}$ where the $\ket i$ are the basis vectors of $\mathbb C^{K-d}$ numbered to start at 0, and define $\ket{\varphi_k}\in \mathbb C^{K-d+1} = \mathbb C^r$ by $\ket{\varphi_k} = \sum_{i=1}^{K-d} \lambda_{k,i-1}\ket{i}$, which are by construction orthogonal to $\ket 0 \in \mathbb C^r$. Then, it is straightforward to check that the set $ Y$ as defined in \eqref{eq:rank-1-PVM-set} is orthonormal. Moreover, it is clear that $\ket{y_k} - V\ket{x_k} = \ket{\eta} \otimes \ket{\varphi_k} \perp \text{range}(V)$. By \eqref{e:isomproperty}, $V^*$ maps such elements to $\mathbf 0$, and hence, $\mathbf 0 = V^*(\ket{y_k} - V\ket{x_k}) = V^*\ket{y_k} - \ket{x_k}$, as required. \hfill $\Box$

\medskip

\textbf{Theorem 3} Let $(R_k)_{k=1}^K$ be a $K$-outcome POVM on a finite-dimensional Hilbert space $ H$. Then, there exists a finite-dimensional Hilbert space $G$ and a $K$-outcome PVM $(P_k)_{k=1}^K$ on $H \otimes G$ such that $R_k = V^*P_k V$ for all $k\in [K]$, where $V: H\to  H \otimes G$ is the isometry given by $V\ket \xi = \ket \xi \otimes \ket 0$.

\medskip

\textit{Proof}. Let $R_k = \sum_{l=1}^{L}\ket{x_{k,l}}\bra{x_{k,l}}$ for non-zero vectors $\ket{x_{k,l}}\in  H$. (This is possible for any choice of $L$ greater than or equal to $\max_k \text{Rank}(R_k)$). Then, $\sum_{k=1}^K\sum_{l=1}^L \ket{x_{k,l}}\bra{x_{k,l}} = I_{ H}$. Applying Lemma~2, we get a Hilbert space $ G = \mathbb C^{KL-d+1}$ and an orthonormal set $ Y = \{\ket{y_{k,l}}:k\in [K],l\in [L]\}\subseteq  H \otimes G$. Let $P^{\perp}\in  B( H\otimes G)$ be the projection onto $ Y^{\perp}$. Define for each $k\in [K]$, \begin{equation}
		P_k = \delta_{k,K} P^{\perp} + \sum_{l=1}^L \ket{y_{k,l}}\bra{y_{k,l}}.
	\end{equation}
By Lemma~2, we also know that $\ket{x_{k,l}}\bra{x_{k,l}} = V^*\ket{y_{k,l}}\bra{y_{k,l}}V$ for all $k\in [K]$ and $l\in [L]$, from which it is immediate that $R_k = V^*P_kV$ for all $k\in [K-1]$. To show that it also holds for the final $k=K$ case, we need to show that the term $V^* P^\perp V$ vanishes, which can be seen as follows: for any $\ket \xi\in H$,
 \begin{align*}
 V^*P^\perp V \ket \xi 
 &= V^* \left (I_{H\otimes G}-\sum_{k.l}\ket {y_{k,l}}\bra{y_{k,l}}\right)\left(\ket\xi \otimes \ket 0\right)\\
 &= V^*\left(\ket \xi\otimes \ket 0 - \sum_{k,l}\braket{x_{k,l}|\xi}\big(\ket{x_{k,l}}\otimes \ket 0 + \ket \eta \otimes \ket {\varphi_{k,l}}\big)\right)\\
 &= V^*\left(- \sum_{k,l}\braket{x_{k,l}|\xi}\ket\eta \otimes  \ket {\varphi_{k,l}}\right) = \mathbf 0,
 \end{align*} where the final equality holds by \eqref{e:isomproperty}. \hfill$\Box$

\medskip

Theorem 3 allows us to replace a POVM on the state $\ket \xi$ with a PVM on the state $\ket\xi \otimes \ket 0$ that yields the same outcome probabilities. We can furthermore iterate the construction in Theorem 3 above to construct an isometry when there are two or more measurement settings. For instance, suppose $\{E_{x,a}:a\in [K]\}$ is a POVM for settings $x\in \{0,1\}$. Then, for the setting $x=0$ applying Theorem~3 we get an isometry $V_0:H\to H\otimes K_0$ and a $K$-outcome PVM $\{P_{0,a}:a\in [K]\}$ on $H\otimes K_0$ such that $E_{0,a} = V_0^*P_{0,a}V_0$ for all $a\in [K]$. We thus can construct an intermediate set of POVMs on the space $H\otimes K$: $\{E_{x,a}^{\prime}:a\in [K]\}$ for $x\in\{0,1\}$, where \begin{equation}
    E_{x,a}^{\prime} = \begin{cases}
    P_{0,a} &\text{ if } x = 0, \\
    \begin{cases}
        V_0E_{1,1}V_0^* + (I - V_0V_0^*) &\text{ if } a = 1 \\
        V_0E_{1,a}V_0^*   &\text{ if } a \geq 2 \\
    \end{cases} &\text{ if } x =1.
    \end{cases} 
\end{equation} The POVM corresponding to the setting $x=1$ may not projective, so we now apply Theorem~3 to the set $\{E_{1,a}^{\prime}:a\in [K]\}$, to get another isometry $V_1:H\otimes K_0 \to (H\otimes K_0)\otimes K_1$ and a PVM $\{P_{1,a}:a\in [K]\}$ such that $E_{1,a}^{\prime} = V_1^*P_{1,a}V_1$ for all $a\in [K]$. Then, we get our final set of POVMs $\{E_{x,a}^{\prime\prime}:a\in [K]\}$ for $x\in \{0,1\}$ given by \begin{equation}
    E_{x,a}^{\prime\prime} = \begin{cases}
    \begin{cases}
        V_1P_{0,1}V_1^* + (I - V_1V_1^*) &\text{ if } a = 1 \\
        V_1P_{0,a}V_1^* &\text{ if } a \geq 2 \\
    \end{cases} &\text{ if } x = 0, \\
    P_{1,a} &\text{ if } x = 1.
    \end{cases} 
\end{equation} Observe that since $(P_{0,a})_{a=1}^K$ is a PVM, so is $(E_{0,a}^{\prime\prime})_{a=1}^K$. Moreover, the final isometry $V_A:H\to H\otimes (K_0\otimes K_1)$ is $V_A = V_1\circ V_0$, which from Theorem~3 is seen to be given by $V_A\ket{x} = \ket{x} \otimes \ket{e_1} \otimes \ket{e_1}$.

\medskip

In the proof of Theorem~1 in SM~\ref{app:Thm-1-proof}, the state is given by $\psi = \ket{\psi}_{AB}\otimes \ket{\psi}_{BC} \otimes \ket{\psi}_{AC}$. We want to replace Alice and Charlie's POVMs with PVMs, for which, we use the procedure mentioned in the previous paragraph. We then get PVMs in a possibly bigger space $H_A \otimes K_A$ (for Alice) and $H_B\otimes K_C$ (for Charlie) and isometries $V_A:H_A \to H_A \otimes K_A$ and $V_C:H_A \to H_C\otimes K_C$ given by $V_A\ket{\xi}  = \ket{\xi} \otimes \ket{0}_A$ (for Alice) and similarly  $V_C\ket{\eta}  = \ket{\eta} \otimes \ket{0}_C$ for Charlie. Then, the new state we are looking at is $(V_A \otimes I_B \otimes V_C)\psi = \ket{0}_A \otimes \ket{\psi}_{AB}\otimes \ket{\psi}_{BC} \otimes \ket{\psi}_{AC} \otimes \ket{0}_C$, justifying Expression~\eqref{eq:psi-prime} in the proof.

\end{document}